\documentclass[twocolumn]{article}
\usepackage{geometry}
\usepackage{bm}
\usepackage{amsmath}
\usepackage{graphicx}
\usepackage[noend]{algpseudocode}
\usepackage{algorithmicx,algorithm}
\usepackage[numbers,sort]{natbib}
\usepackage[labelfont=bf]{caption}
\usepackage{float}
\usepackage{multicol,lipsum}

\usepackage[T1]{fontenc}
\usepackage[utf8]{inputenc}
\usepackage{authblk}
\usepackage[colorlinks,linkcolor=blue,citecolor=blue,urlcolor=blue]{hyperref}
\usepackage[title]{appendix}
\usepackage{lettrine}
\usepackage{hyperref}

\graphicspath{{figures/}} 

\numberwithin{equation}{section}

\title{\textbf{{\LARGE Identification of brain states, transitions, and communities using functional MRI}}}

\author[a,b,*]{\small Lingbin Bian }
\author[a]{\small Tiangang Cui }
\author[c]{\small B.T. Thomas Yeo}
\author[b]{\small Alex Fornito }
\author[b,d,+,*]{\small Adeel Razi }
\author[a,+]{\small Jonathan Keith }

\affil[a]{School of Mathematics, Monash University, Australia}
\affil[b]{Turner Institute for Brain and Mental Health, School of Psychological Sciences and Monash Biomedical Imaging, Monash University, Australia}
\affil[c]{Department of Electrical and Computer Engineering, National University of Singapore, Singapore}
\affil[d]{Wellcome Centre for Human Neuroimaging, University College London, United Kingdom}
\affil[+]{Joint senior authors}
\affil[*]{Corresponding authors: Lingbin Bian (lingbin.bian@monash.edu) and Adeel Razi (adeel.razi@monash.edu)}
\date{} 

\begin{document}
\renewcommand{\figurename}{\textbf{Fig.}} 
\maketitle



\begin{abstract}
\setlength{\parindent}{0pt} \setlength{\parskip}{1.5ex plus 0.5ex
minus 0.2ex} 
Brain function relies on a precisely coordinated and dynamic balance between the functional integration and segregation of distinct neural systems. Characterizing the way in which neural systems reconfigure their interactions to give rise to distinct but hidden brain states remains an open challenge. In this paper, we propose a Bayesian model-based characterization of latent brain states and showcase a novel method based on posterior predictive discrepancy using the latent block model to detect transitions between latent brain states in blood oxygen level-dependent (BOLD) time series. The set of estimated parameters in the model includes a latent label vector that assigns network nodes to communities, and also block model parameters that reflect the weighted connectivity within and between communities. Besides extensive \textit{in-silico} model evaluation, we also provide empirical validation (and replication) using the Human Connectome Project (HCP) dataset of 100 healthy adults. Our results obtained through an analysis of task-fMRI data during working memory performance show appropriate lags between external task demands and change-points between brain states, with distinctive community patterns distinguishing fixation, low-demand and high-demand task conditions.
\end{abstract}

\section*{ }
\lettrine[lines=3]{\textbf{I}}{ \textbf{identifying}} 
changes in brain connectivity over time can provide insight into fundamental properties of human brain dynamics.  However, the definition of discrete brain states and the method of identifying the states has not been commonly agreed  \citep*{Kringelbach2020}. Experiments targeting unconstrained spontaneous `resting-state' neural dynamics \citep*{Lurie2020, Razi2016, Razi2017, Razi2015, Friston2014, Friston2020, Allen2014, Hutchison2013, Calhoun2014} have limited ability to infer latent brain states or determine how the brain segues from one state to another because it is not clear whether changes in brain connectivity are induced by variations in neural activity (for example induced by cognitive or vigilance states) or fluctuations in non-neuronal noise \citep*{Power2017, Parkes2018, Aquino2020}. A recent study with naturalistic movie stimuli used a hidden Markov model to explore dynamic jumps between discrete brain states and found that the variations in the sensory and narrative properties of the movie can evoke discrete brain processes \citep*{Meer2020}. Task-fMRI studies with more restrictive constraints on stimuli have demonstrated that functional connectivity exhibits variation during motor learning \citep*{Bassett2011} and anxiety-inducing speech preparation \citep*{Cribben2012}. Although task-based fMRI experiments can, to some extent, delineate the external stimuli (for example, the onset and duration of stimuli in the block designed experiments), which constitute reference points against which to identify changes in the observed signal, this information does not precisely determine the timing and duration of the latent brain state relative to psychological processes or neural activity. For example, an emotional stimulus may trigger a neural response which is delayed relative to stimulus onset and which persists for some time after stimulus offset. Moreover, the dynamics of brain states and functional networks are not induced only by external stimuli, but also by unknown intrinsic latent mental processes \citep*{Taghia2018a}. Therefore, the development of noninvasive methods for identifying transitions of latent brain states during both task performance and task-free conditions is necessary for characterizing the spatiotemporal dynamics of brain networks.\par

Change-point detection in multivariate time series is a statistical problem that has clear relevance to identifying transitions in brain states, particularly in the absence of knowledge regarding the experimental design. Several change-point detection methods based on spectral clustering \citep*{Luxburg2007, Cribben2017} and dynamic connectivity regression (DCR) \citep*{Cribben2012} have been previously developed and applied to the study of fMRI time series, and these have enhanced our understanding of brain dynamics. However, change-point detection with spectral clustering only evaluates changes to the component eigenstructures of the networks but neglects the weighted connectivity between nodes, while the  DCR method only focuses on the sparse graph but ignores the modules of the brain networks Other change-point detection strategies include a frequency-specific method \citep*{Ombao2019}, applying a multivariate cumulative sum procedure to detect change-points using EEG data, and methods which focus on large scale network estimation in fMRI time series \citep*{Frick2014, Cho2015, TWang2017, Park2018}. Many fMRI studies use sliding window methods for characterizing the time-varying functional connectivity in time series analysis \citep*{Chang2010, Handwerker2012, Allen2014, Zalesky2014, Monti2014, Jeong2016, Lurie2020}. Methods based on hidden Markov models (HMM) are also widely used to analyze transient brain states \citep*{Vidaurre2016,Vidaurre2017,Vidaurre2018}.

\par
A {\em community} is defined as a collection of nodes that are densely connected in a network. The problem of community detection is a topical area of network science \citep*{Wang2017, Jin2015}. How communities change or how the nodes in a network are assigned to specific communities is an important problem in the characterization of networks. Although many community detection problems in network neuroscience are based on modularity \citep*{Bassett2011, Newman2006, Bassett2013}, recently a hidden Markov stochastic block model combined with a non-overlapping sliding window was applied to infer dynamic functional connectivity for networks, where edge weights were  only binary and the candidate time points evaluated were not consecutive \citep*{Robinson2015, Ting2020}. More general weighted stochastic block models \citep*{Aicher2015} have been used to infer structural connectivity for human lifespan analysis \citep*{Faskowitz2018} and to infer functional connectivity in the mesoscale architecture of drosophila, mouse, rat, macaque, and human connectomes \citep*{Betzel2018}. However, these studies using the weighted stochastic block model only explore the brain network over the whole time course of the experiment and neglect dynamic properties of networks. Weighted stochastic block models \citep*{Aicher2015} are described in terms of exponential families (parameterized probability distributions), with the estimation of parameters performed using variational inference \citep*{Hoffman2013, Blei2017}. Another relevant statistical approach introduces a fully Bayesian latent block model \citep*{Nobile2007, Wyse2012}, which includes both a binary latent block model and a Gaussian latent block model as special cases. The Gaussian latent block model is similar to the weighted stochastic block model, but different methods have been used for parameter estimation, including Markov chain Monte Carlo (MCMC) sampling.

Although there is a broad literature exploring change-point detection, and also many papers that discuss community detection, relatively few papers combine these approaches, particularly from a Bayesian perspective. In this paper, we develop Bayesian model-based methods which unify change-point detection and community detection to explore when and how the community structure of discrete brain state changes under different external task demands at different time points using functional MRI. 
There are several advantages of our approach compared to existing change-point detection methods. Compared to data-driven methods like spectral clustering \citep*{Luxburg2007, Cribben2017} and DCR \citep*{Cribben2012}, which either ignore characterizing the weighted connectivity or the community patterns, the fully Bayesian framework and Markov chain Monte Carlo method provide flexible and powerful strategies that have been under-used for characterizing the latent properties of brain networks, including the dynamics of both the community memberships and weighted connectivity properties of the nodal community structures. Existing change-point detection methods based on the stochastic block model all use non-overlapping sliding windows and were applied only to binary brain networks \citep*{Robinson2015, Ting2020}. In contrast to the stochastic block model used in time-varying functional connectivity studies, the Gaussian latent block model used in our work considers the correlation matrix as an observation without imposing any arbitrary thresholds, so that all the information contained in the time series is preserved, resulting in more accurate detection of change-points. Moreover, unlike methods based on fixed community memberships over the time course \citep*{Ting2020}, our methods consider both the community memberships and parameters related to the weighted connectivity to be time varying, which results in more flexible estimation of both community structure and connectivity patterns. Furthermore, our Bayesian change-point detection method uses overlapping sliding windows that assess {\em all} of the potential candidate change-points over the time course, which increases the resolution of the detected change-points compared to methods using non-overlapping windows \citep*{Robinson2015, Ting2020}. Finally, the proposed Bayesian change-point detection method is computationally efficient, scaling to whole-brain networks potentially covering hundreds of nodes within a reasonable time frame in the order of tens of minutes. 
\par

\begin{figure*}[!ht]
\centering
\includegraphics[width=0.65\linewidth]{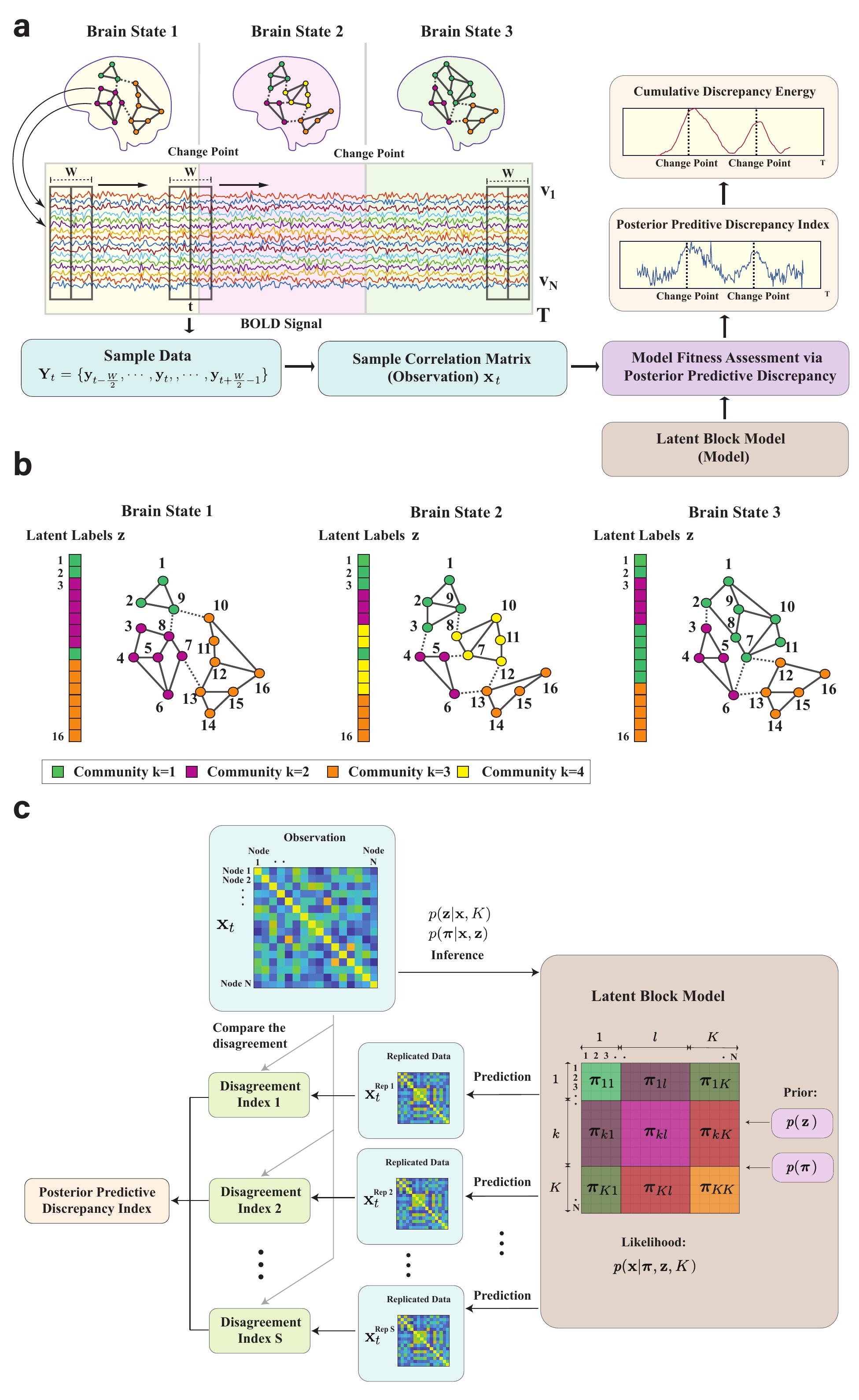}
\caption{\footnotesize The framework for identifying brain states, transitions and communities: \textbf{a} Schematic of the proposed Bayesian change-point detection method. Three different background colors represent three brain states of individual subjects with different community architectures. The colors of the nodes represent community memberships. A sliding window of width $W$ centered at $t$ is applied to the time series. The different colored time series correspond to BOLD time series for each node. The sample correlation matrix $\textbf{x}_{t}$ (i.e., an observation for our Bayesian model) is calculated from the sample data $\textbf{Y}_{t}$ within the sliding window. We use the Gaussian latent block model to fit the observations and evaluate model fits to the observations to obtain the posterior predictive discrepancy index (PPDI). We then calculate the cumulative discrepancy energy (CDE) from the PPDI and use the CDE as a scoring criterion to estimate the change-points of the community architectures. \textbf{b} Dynamic community memberships of networks with $N=16$ nodes. A latent label vector $\textbf{z}$ contains the labels ($k$) of specific communities for the nodes. Nodes of the same color are located in the same community. The dashed lines represent the (weighted) connectivity between communities and the solid lines represent the (weighted) connectivity within the communities. \textbf{c} Model fitness assessment. The observation is the realized adjacency matrix; different colors in the latent block model represent different blocks with the diagonal blocks representing the connectivity within a community and the off-diagonal blocks representing the connectivity between communities. To demonstrate distinct blocks of the latent block model, in this schematic we group the nodes in the same community adjacently and the communities are sorted. In reality, the labels of the nodes are mixed with respect to an adjacency matrix. The term $\bm{\pi}_{kl}$ represents the model parameters in block $kl$.}
\label{Figure_1_schematic}
\end{figure*}

Our paper presents four main contributions, namely: (i) we quantitatively characterize discrete brain states with weighted connectivity and time-dependent community memberships, using the latent block model within a temporal interval between two consecutive change-points; (ii) we propose a new Bayesian change-point detection method called \textit{posterior predictive discrepancy} (PPD) to estimate transition locations between brain states, using a Bayesian model fitness assessment; (iii) in addition to the locations of change-points, we also infer the community architecture of discrete brain states, which we show are distinctive of 2-back, 0-back, and fixation conditions in a working-memory task-based fMRI experiment, and; (iv) we further empirically find that the estimated change-points between brain states show appropriate lags compared to the external working memory task conditions.

\begin{figure*}[!ht]
\centering
\includegraphics[width=0.9\linewidth]{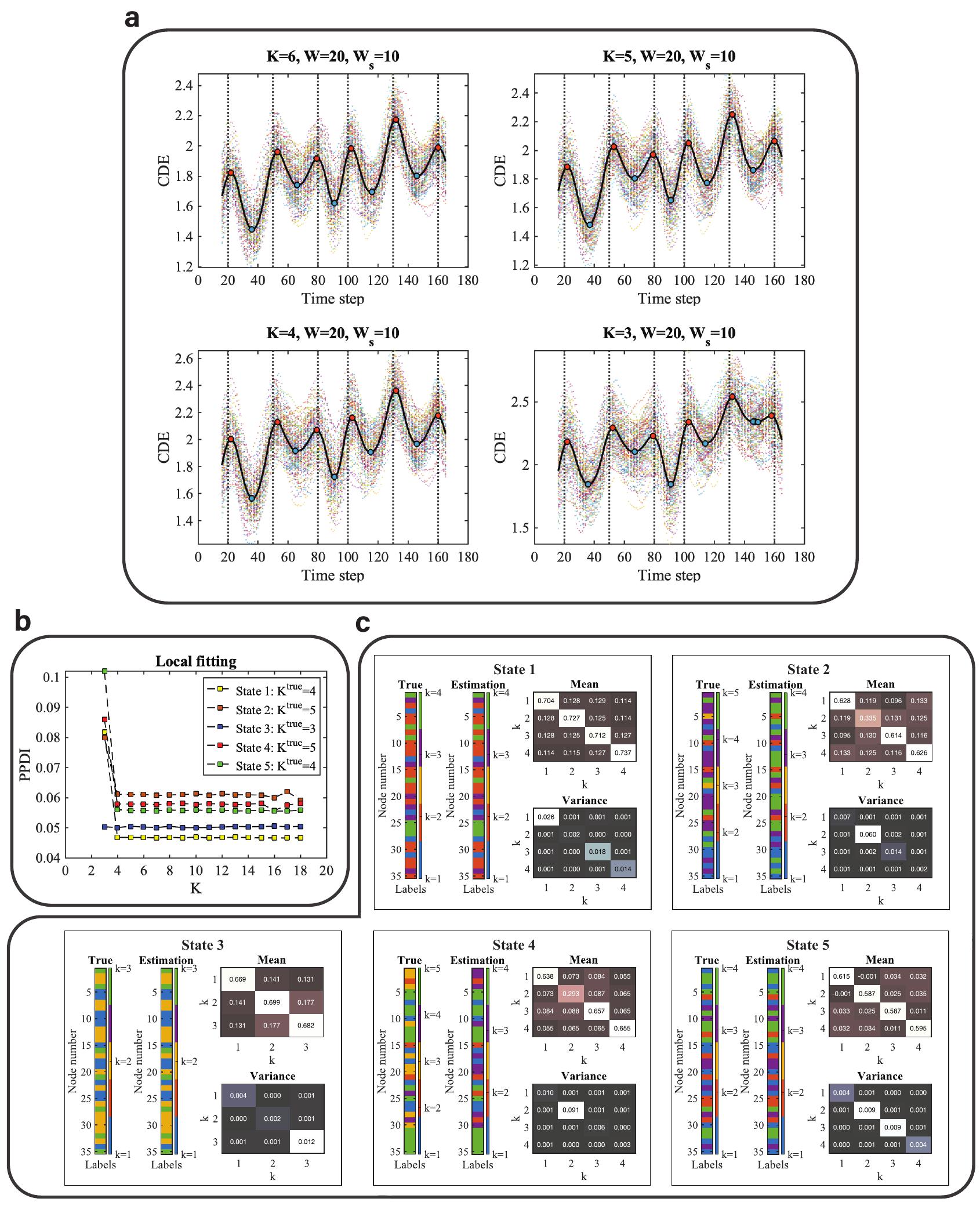}
\caption{\footnotesize Results of method validation using synthetic data: \textbf{a} CDE of the multivariate Gaussian data with SNR=5dB using different models ($K$=6, 5, 4, and 3). The sliding window size for converting from time series to correlation matrices sequence is $W=20$, whereas (for smoothing) the sliding window size for converting from PPDI to CDE is $W_{s}=10$. The vertical dashed lines are the locations of the true change-points ($t$=20, 50, 80, 100, 130, and 160). The colored scatterplots in the figures are the CDEs of individual (virtual) subjects and the black curve is the group CDE (averaged CDE over 100 subjects). The red points are the local maxima and the blue points are the local minima. \textbf{b} Local fitting with different models (from $K$=3 to 18) for synthetic data (SNR=5dB). Different colors represent the PPDI values of different states with the true number of communities $K^{true}$. \textbf{c} The estimation of community constituents for SNR=5dB at each discrete state: $t$=36, 66, 91, 116, 146) for brain states 1 to 5, respectively. The estimations of the latent label vectors (\textbf{Estimation}) and the label vectors (\textbf{True}) that determine the covariance matrix in the generative model are shown as bar graphs. The strength and variation of the connectivity within and between communities are represented by the block mean and variance matrices within each panel.}
\label{Figure_2_synthetic}
\end{figure*}

\section*{Results}

Our proposed method is capable of identifying transitions between discrete brain states and infer the patterns of connectivity between brain regions that underlie those brain states by modelling time-varying dynamics in BOLD signal under different stimuli. In this section, we validate our proposed methodology by applying Bayesian change-point detection and network estimation to both synthetic data and real fMRI data. The Bayesian change-point detection method is described in Fig. \ref{Figure_1_schematic} and the mathematical formulation and detailed descriptions are in the \textbf{Methods} section (also see \textbf{Supplementary information}). We first use synthetic multivariate Gaussian data for extensive validation and critically evaluate the performance of our change-point detection and sampling algorithms. For real data analysis, we use working memory task fMRI (WM-tfMRI) data from the Human Connectome Project (HCP) \citep*{Barch2013}. We extracted the time series of 35 nodes whose MNI coordinates were determined by 
significant activations obtained via clusterwise inference using FSL \citep*{smith2004}.

\subsection*{Method validation using synthetic data}

To validate the Bayesian change-point detection algorithm, we first use synthetic data with the signal to noise ratio (SNR)=5dB. The simulated states of segments between two true change-points in the synthetic data could be repeating or all different, depending on the settings of the parameters in the generative model. A detailed description of the generative model and parameter settings for simulating the synthetic data are provided in \textbf{Supplementary Section 1}. Further simulation results with different levels of SNR (SNR=10dB, SNR=0dB, and SNR=-5dB) are provided in \textbf{Supplementary Figures 1, 2, and 3}. The resulting group-level cumulative discrepancy energy (CDE) scores (see \textbf{Methods} section for how we define CDE) using models with different values of $K$, where $K$ is the number of communities, are shown in Fig. \ref{Figure_2_synthetic}a. We use a latent block model to fit the adjacency matrix at consecutive time points for change-point detection, which we call global fitting. The local maxima (red dots) of the group CDE indicate locations of change-points and the local minima (blue dots) correspond to distinct states that differ in their community architecture. We find that the local maxima (red dots) are located very close to the true change-points in all of the graphs (in Fig. \ref{Figure_2_synthetic}a) which means that the global fitting has good performance. Here we clarify that global fitting is used to estimate the locations of the change-points or transitions of brain states, and local fitting is used to select a latent block model to estimate the community structures of discrete brain states (see \textbf{Methods} section for a detailed explanation of global and local fitting).

Next, using the global fitting results, with $K=6$ and $W=20$, where $W$ is the width of the sliding (rectangular) window, we find the local minima (the blue dots) locations to be $t=\lbrace 36, 66, 91, 116, 146 \rbrace$, where each location corresponds to a discrete state. Next, we use local fitting to select a model (i.e. $K$ for local inference) to infer the community membership and model parameters relating to the connectivity of the discrete states. For local inference, the group averaged adjacency matrix is considered as the observation. We assess the goodness of fit between observation and a latent block model with various values of $K$ (from $K=3,\cdots,18$) using posterior predictive discrepancy for each local minimum, as shown in Fig. \ref{Figure_2_synthetic}b. We selected the value of $K$ at which the curve starts to flatten as the preferred model. We find that the model assessment curves for states 1, 2, 4, and 5 flatten at $K=4$, whereas the model assessment curve for state 3 is flat over the entire range (from $K=3$ and up). Therefore the selected models are $K=\lbrace 4,4,3,4,4 \rbrace$ for states 1 to 5, respectively.

To validate the MCMC sampling of the density $p(\textbf{z}|\textbf{x},K)$, we compare the estimate of the latent label vector to the ground truth of the node memberships. Fig. \ref{Figure_2_synthetic}c shows the inferred community architectures of the discrete states including the estimated latent label vectors and the model parameters of block mean and variance. The true label vectors that determine the covariance matrix in the generative models are also included in this figure. We use the most frequent latent label vectors in the Markov chain after the burn-in steps as the estimate. Note that  {\em label-switching} occurs in the MCMC sampling, which is a well-known problem in Bayesian estimation of mixture models \citep*{Stephens2000}. In the results presented here, the node memberships have been relabelled to correct for label switching. The algorithm used for this purpose is described in  \textbf{Supplementary Section 2}. We find that the estimated latent label vectors are (largely) consistent with the ground truth of labels that determined the covariance matrix. The discrepant `True' and `Estimation' patterns with respect to states 2 and 4 are due to the bias induced by the selected model ($K=5$ for the ground truth and $K=4$ for the selected model). Although the colors of the labels in the `True' and `Estimation' patterns are discrepant, we can see that the values of the labels are largely consistent, with some labels of $k=5$ missing in the `Estimation' pattern compared to the `True' pattern.

Given the estimated latent label vector, we then draw samples of the block mean and variance from the posterior $p(\bm{\pi|\textbf{x},\textbf{z}})$ conditional on the estimated latent label vector $\textbf{z}$. However, there is no ground truth for the block mean and variance when we generate the synthetic data. The validation of sampling model parameters is illustrated in the \textbf{Supplementary Figure 4}.

\subsection*{Method validation using working memory (WM) task-fMRI data}

\begin{table*}[h!]
\centering
\begin{tabular}{lllllllll}

  &  &\multicolumn{3}{c}{MNI coordinates} &\multicolumn{3}{c}{Voxel locations}\\

Node number& Z-MAX& x& y& z& x& y& z &Region name\\
\hline
1&		4.97&	48&	-58&	22&		21&	34&	47&		Angular Gyrus\\
2&		9.61	&	36&	8&	12&		27&	67&	42&		Central Opercular Cortex\\
3&		8.27&	-36&	4&	12&		63&	65&	42&		Central Opercular Cortex\\
4&		6.48&	40&	34&	-14&		25&	80&	29&		Frontal Orbital Cortex\\
5&		7.83	&	-12&	46&	46&		51&	86&	59&		Frontal Pole\\
6&		4.84	&	54&	32&	-4&		18&	79&	34&		Inferior Frontal Gyrus\\
7&		6&		52&	38&	10&		19&	82&	41&		Inferior Frontal Gyrus\\
8&		4.38	&	-52&	40&	6&		71&	83&	39&		Inferior Frontal Gyrus\\
9&		6.05	&	52&	-70&	36&		19&	28&	54&		Inferior Parietal Lobule PGp R\\
10&		7.26	&	-48&	-68&	34&		69&	29&	53&		Inferior Parietal Lobule PGp L\\
11&		6.18	&	44&	-24&	-20&		23&	51&	26&		Inferior Temporal Gyrus\\
12&		9.54	&	36&	-86&	16&		27&	20&	44&		Lateral Occipital Cortex\\
13&		8.04	&	-30&	-80&	-34&		60&	23&	19&		Left Crus I\\
14&		7.6&		-8&	-58&	-52&		49&	34&	10&		Left IX\\
15&		6.9&		-22&	-48&	-52&		56&	39&	10&		Left VIIIb\\
16&		14.5	&	6&	-90&	-10&		42&	18&	31&		Lingual Gyrus\\
17&		10.3	&	30&	10&	58&		30&	68&	65&		Middle Frontal Gyrus\\
18&		6.61	&	66&	-30&	-12&		12&	48&	30&		Middle Temporal Gyrus\\
19&		4.53	&	-68&	-34&	-4&		79&	46&	34&		Middle Temporal Gyrus\\
20&		14.5	&	18&	-88&	-8&		36&	19&	32&		Occipital Fusiform Gyrus\\
21&		5.06	&	-12&	-92&	-2&		51&	17&	35&		Occipital Pole\\
22&		9.87	&	6&	40&	-6&		42&	83&	33&		Paracingulate Gyrus\\
23&		12&		42&	-16&	-2&		24&	55&	35&		Planum Polare\\
24&		11.3&	-40&	-22&	0&		65&	52&	36&		Planum Polare\\
25&		9.03&	38&	-26&	66&		26&	50&	69&		Postcentral Gyrus\\
26&		8.31	&	-10&	-60&	14&		50&	33&	43&		Precuneus Cortex\\
27&		5.7&		46&	-60&	-42&		22&	33&	15&		Right Crus I\\
28&		8.34	&	32&	-80&	-34&		29&	23&	19&		Right Crus I\\
29&		10.9	&	32&	-58&	-34&		29&	34&	19&		Right Crus I\\
30&		6.41	&	10&	-8&	-14&		40&	59&	29&		Right Hippocampus\\
31&		6.19	&	32&	-52&	2&		29&	37&	37&		Right Lateral Ventricle\\
32&		7.69	&	24&	-46&	16&		33&	40&	44&		Right Lateral Ventricle\\
33&		6.13	&	0&	10&	-14&		45&	68&	29&		Subcallosal Cortex\\
34&		10.7	&	48&	-44&	46&		21&	41&	59&		Supramarginal Gyrus\\ 
35&		4.23&	-50&	-46&	10&		70&	40&	41&		Supramarginal Gyrus\\ 

\hline
\end{tabular}
\caption{\footnotesize Significant activations of cluster wise inference (cluster-corrected Z>3.1, P<0.05): Activations are described in terms of local maximum Z (Z-MAX) statistic within each cluster including the activations of all contrast maps among 2-back, 0-back, and fixation.}
\label{table:1}
\end{table*}

In this analysis, we used preprocessed working memory (WM)-tfMRI data obtained from 100 unrelated healthy adult subjects under a block designed paradigm, available from the Human Connectome Project (HCP) \citep*{Barch2013}. We mainly focused on the working memory load contrasts of 2-back vs fixation, 0-back vs fixation, or 2-back vs 0-back, and determine the brain regions of interest from the GLM analysis. After group-level GLM analysis, we obtained cluster activations with locally maximum Z statistics for different contrasts. The results in the form of thresholded local maximum Z statistic (Z-MAX) maps are shown in \textbf{Supplementary Figure 5}. The light box views of thresholded local maximum Z statistic with different contrasts are provided in \textbf{Supplementary Figures 6}. Significant activations obtained by clusterwise inference and the corresponding MNI coordinates with region names are shown in Table \ref{table:1}. We finally extracted the time series of 35 brain regions corresponding to the MNI coordinates. Refer to \textbf{Methods} section for the details of experimental design, GLM analysis and time series extraction. 

\begin{figure*}[!ht]
\centering
\includegraphics[width=0.8\linewidth]{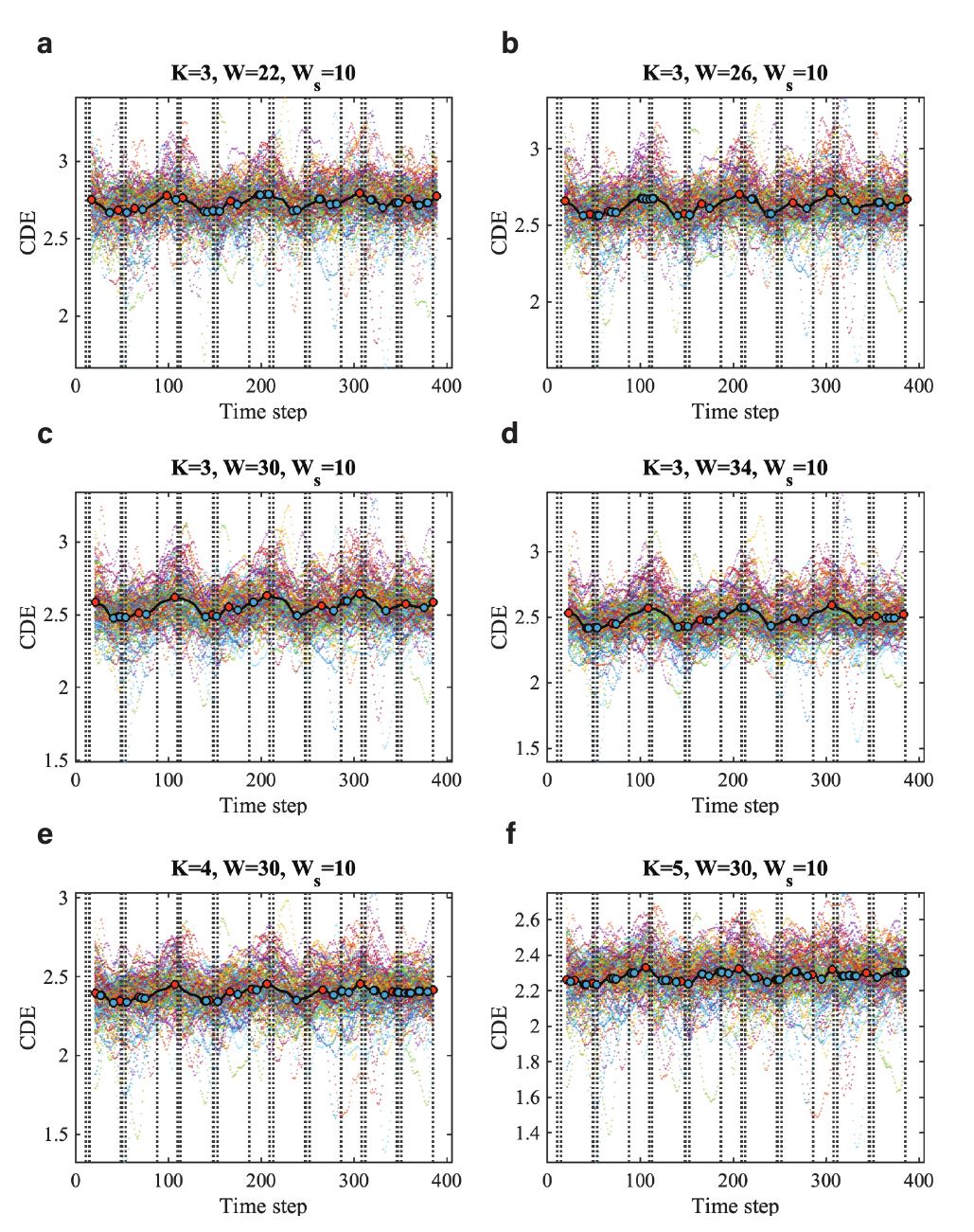}
\caption{\footnotesize The results of Bayesian change-point detection for working memory tfMRI data (session 1, LR): Cumulative discrepancy energy (CDE) with different sliding window sizes ($W$=22, 26, 30, and 34; \textbf{a}-\textbf{d} under the model $K=3$) and different models (K=3, 4, and 5; \textbf{c}, \textbf{e} and \textbf{f} using a sliding window of $W=30$). $W_{s}$ is width of the sliding window used for converting from PPDI to CDE. The vertical dashed lines are the times of onset of the stimuli, which are provided in the EV.txt files in the released data. The multi-color scatterplots in the figures represent the CDEs of individual subjects and the black curve is the group CDE (averaged CDE over 100 subjects). The red dots are the local maxima, which are taken to be the locations of change-points, and the blue dots are the local minima, which are used for local inference of the discrete brain states.}
\label{Figure_3_changepoint}
\end{figure*}

\begin{figure*}[!ht]
\centering
\includegraphics[width=0.8\linewidth]{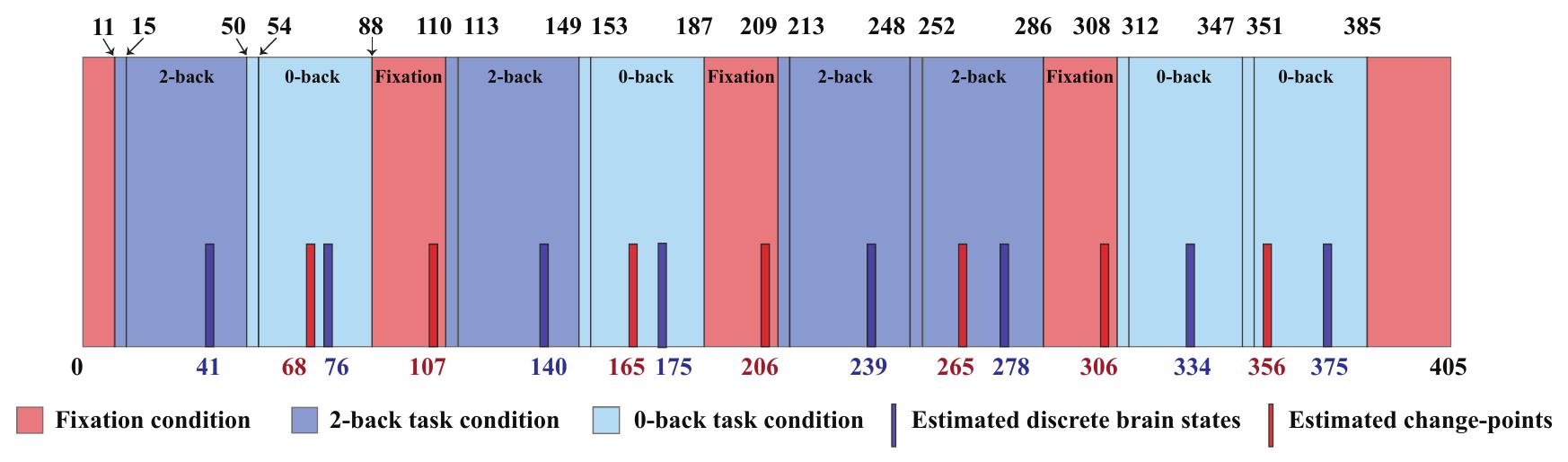}
\caption{\footnotesize Detected change-points and the locations of brain states matching the task blocks for working memory tfMRI data (session 1, LR) with $K=3$, and $W=30$. The numbers in the small rectangular frame are the boundaries of the external task demands, the background colors in the large rectangular are the different task conditions, and the blue and red bars with specified numbers are the estimated locations of discrete brain states and change-points.}
\label{Figure_4_esti_changepoint}
\end{figure*}

\subsubsection*{Change-point detection for tfMRI time series}
In the main text, we illustrate the results using the HCP working memory data of session 1 i.e. with the polarity of Left to Right (LR). The replication of results obtained by using session 2 (RL) are demonstrated in \textbf{Supplementary Figures 10 to 15)} and \textbf{Supplementary Table 1}. We compare the brain states of different working memory loads for a specific kind of picture (tool, body, face, and place) involved in the experiments. As there is no repetition of task conditions in a single session, the estimated patterns of brain states do not recur in LR session. One can compare the LR and RL session for the recurrence of a specific task condition. To detect change-points in the extracted time series, we first converted each time series into a sequence of correlation matrices for each subject. We then modeled this sequence of correlation matrices for each subject using the latent block model and evaluated posterior predictive discrepancy (PPD) to assess the model fitness. Next, we converted the resulting PPD index (PPDI) to a CDE score for each subject. For group-level analysis, we averaged the resulting individual CDE scores over 100 subjects to obtain a sequence of group CDE as shown in Fig. \ref{Figure_3_changepoint} with different window sizes $W$=22, 26, 30, 34 (Fig. \ref{Figure_3_changepoint}a-d). We chose the window size for converting from PPDI to CDE to be a constant $W_{s}=10$ for all of the assessments. The multi-colored scatterplots in the figures are the individual CDE scores. Although there are some false positives in terms of both local maxima and local minima (here false positives are defined as multiple points of local minima or local maxima that should be discarded in a single task block), we note that the onsets of the stimuli precede the inferred local maxima, and the local minima also show appropriate lags (for example, about 10 frames, or 7 seconds as shown in Fig. \ref{Figure_3_changepoint}c) compared to the mid-points of the working memory blocks. For fixation blocks, the local maxima show lags compared to the mid-points of the blocks.
These lags are likely due to the delay in the haemodynamic response of brain activation. With the same number of communities $K=3$, we found there are more false positives with window size $W=22$ compared to $W=26$, $W=30$ and $W=34$. This is because there are fewer sample data contained in the sliding window if the window size is smaller. We also tried different models with $K=4$ and $K=5$. We found that there are more false positives with larger values of $K$. Larger values of $K$ imply more blocks in the model, which gives rise to relatively better model fitness. In this situation, there will be less distinction between relatively static brain states and transition states with change-points in the window.
The false positives among the local minima and local maxima are also influenced by the window size $W_{s}$ used for transforming from PPDI to CDE. A larger window size (for example $W_{s}=30$)  reduces the accuracy of the estimates and results in false negatives. Too small a value of $W_{s}$ increases the false positive rate. We found that $W_{s}=10$ works well for all of the real data analyses. The time spent to run the posterior predictive assessment on each subject ($T$=405 frames, posterior predictive replication number $S$=50, $K$=3, and the window size $W$=30) by using a 2.6 GHz Intel Core i7 processor unit was about 10 minutes.

\begin{figure}[!ht]
\centering
\includegraphics[width=1.0\linewidth]{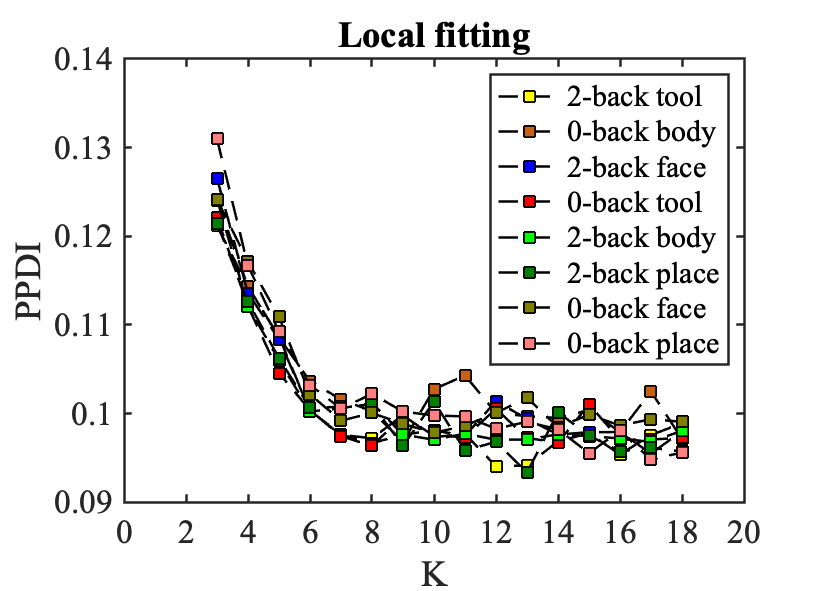}
\caption{\footnotesize Local fitting between averaged adjacency matrix and models from $K$=3 to 18. Different colors represent the PPDI values of different brain states.}
\label{Figure_5_real_localfitting}
\end{figure}

\begin{figure*}[!ht]
\centering
\includegraphics[width=0.8\linewidth]{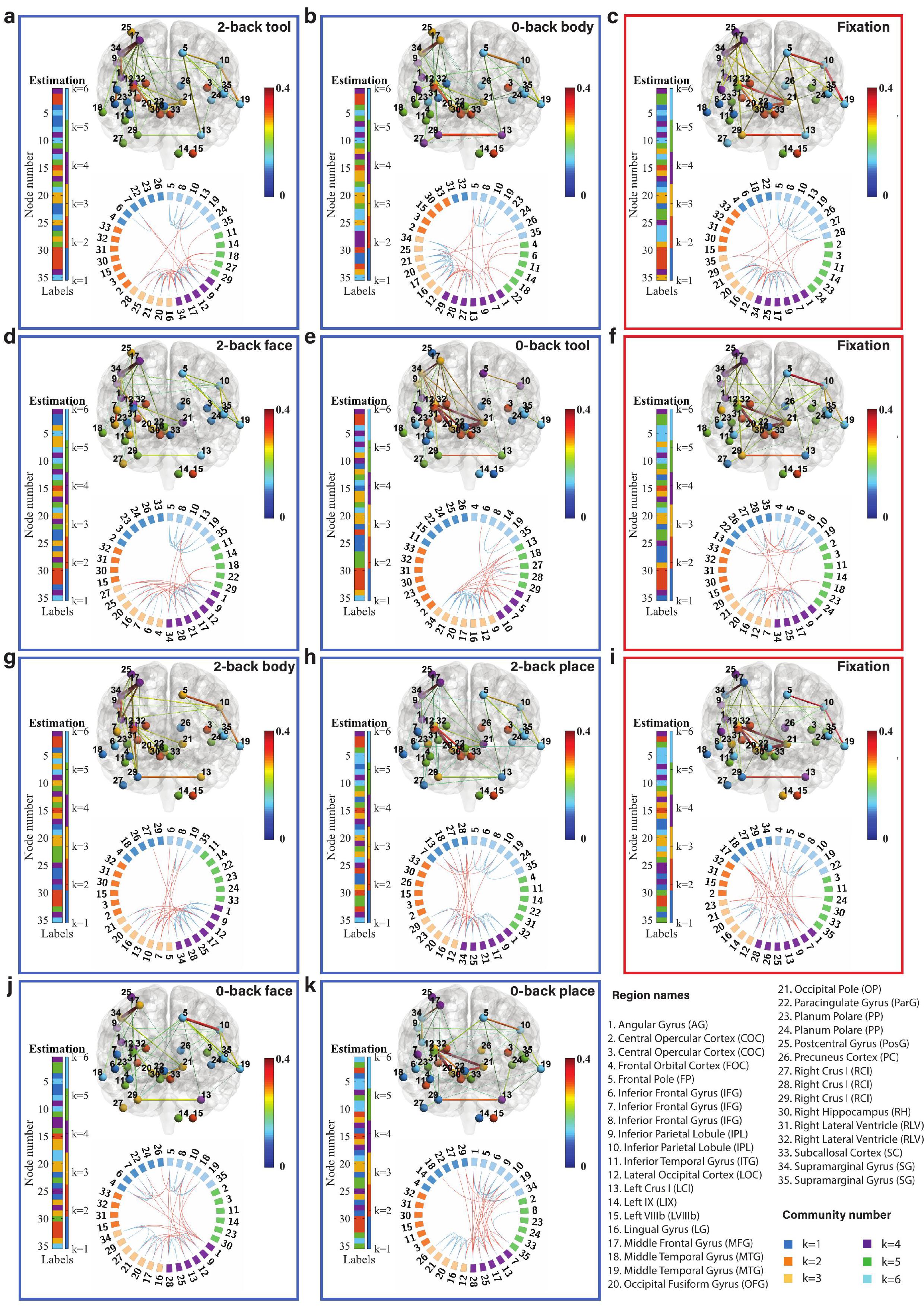}
\caption{\footnotesize Community structure of the discrete brain states: The figures with blue frames represent brain states corresponding to working memory tasks (2-back tool at $t=41$; 0-back body at $t=76$; 2-back face at $t=140$; 0-back tool at $t=175$; 2-back body at $t=239$; 2-back place at $t=278$; 0-back face at $t=334$; and 0-back place at $t=375$ in \textbf{a}-\textbf{k}) and those with red frames represent brain states corresponding to fixation (fixation at $t$=107, 206, and 306 in \textbf{c}, \textbf{f}, and \textbf{i}). Each brain state shows connectivity at a sparsity level of 10\%. The different colors of the labels represent community memberships. The strength of the connectivity is represented by the colors shown in the bar at the right of each frame. In Circos maps, nodes in the same community are adjacent and have the same color. Node numbers and abbreviations of the corresponding brain regions are shown around the circles. In each frame, different colors represent different community numbers. The connectivity above the sparsity level is represented by arcs. The blue links represent connectivity within communities and the red links represent connectivity between communities. 
}
\label{Figure_6_brain_states}
\end{figure*}

The `local inference' is defined as a way to estimate the discrete brain state corresponding to each task condition via Bayesian modelling. The group averaged dynamic functional networks were analyzed by performing `local inference' as follows. In this experiment, we used results obtained for $K=3$ and $W=30$ (see Fig. \ref{Figure_3_changepoint}c). We first listed all of the local maxima and minima, where the time points with distances smaller than 8 are grouped as vectors. Maxima and minima deemed to be false positives were discarded. The time points corresponding to the local minimum value of group CDE are (41, 46, 48, 50, 54) and (140, 147, 152). These were determined as single time points corresponding to discrete brain states, specifically 41 and 140 respectively, with all the other elements in the vectors presumed to be false positives and discarded. Time points with CDE value difference smaller than 0.002 were also discarded (points (191, 192) and (290, 292)). Then the resulting estimated change-point locations (maxima) are at $\lbrace 68, 107, 165, 206, 265, 306, 356 \rbrace$, and the estimated time points of the discrete brain states (minima) are $\lbrace41, 76, 140, 175, 239, 278, 334, 375\rbrace$. A comparison of the detected change-points to the task blocks for working memory tfMRI data are shown in Fig. \ref{Figure_4_esti_changepoint}.

\subsubsection*{Local inference for discrete brain states}
For `local inference', we first calculated the group averaged adjacency matrix with a window of $W_{g}=20$, for all brain states. The center of the window is located at the time point of the local minimum value. We evaluated the goodness of fit for models with different values of $K$ (Fig. \ref{Figure_5_real_localfitting}). The results demonstrate that the goodness of fit trends to flat at $K=6$. To avoid empty communities, $K=6$ is then selected as the number of communities in local inference.  Note that the value of $K$ is unchanged in Markov chain Monte Carlo estimation, but an empty community containing no labels may take place. In the remainder of this section, we used the model with $K=6$ for all brain states. The times spent to run the estimation for latent label vector and model parameters for a single discrete brain state (MCMC sampling number $S_{s}$=200, $K$=6, and the window size $W$=20) by using a 2.6 GHz Intel Core i7 processor unit were about 1.85 and 1.25 seconds respectively.

The inferred community structures are visualized using BrainNet Viewer \citep{Xia2013} and Circos maps \citep{Krzywinski2009} as shown in Fig. \ref{Figure_6_brain_states}. Estimated latent label vectors are visualized using different colors to represent different communities. The nodes are connected by weighted links at a sparsity level of 10\% (we also visualized the brain states with sparsity levels of 20\% and 30\%: \textbf{Supplementary Figures 7 and 8}). The density and variation of connectivity within and between communities are characterized by the estimated block mean matrix and block variance matrix in \textbf{Supplementary Figure 9 and 15}.
We first describe the working memory tasks involving the 2-back tool (Fig. \ref{Figure_6_brain_states}a), 0-back tool (Fig. \ref{Figure_6_brain_states}e), and fixation (Fig. \ref{Figure_6_brain_states}c, f, i). The locations of fixation states are considered as the locations of the change-points at 107, 206, and 306 (we consider the fixation state as a transition buffer between two working memory blocks). We found that the connectivity between the inferior parietal lobule (node 9) and middle frontal gyrus (node 17), and the connectivity between the inferior parietal lobule (node 9) and supramarginal gyrus (node 34) are increased significantly both in 2-back and 0-back working memory compared to fixation. 

For 2-back face (Fig. \ref{Figure_6_brain_states}d) and 0-back face (Fig. \ref{Figure_6_brain_states}j), The connectivity between inferior parietal lobule (node 9) and supramarginal gyrus (node 34) and the connectivity between angular gyrus (node 1) and supramarginal gyrus (node 34) are increased in 2-back compared to 0-back and fixation. There is reduced connectivity between the  lateral occipital cortex (node 12), occipital fusiform gyrus (node 20), and occipital pole (node 21) in 2-back and 0-back compared to fixation. There is reduced connectivity between the frontal pole (node 5) and inferior parietal lobule (node 10) only in 2-back. 

For task blocks with body parts pictures (Fig. \ref{Figure_6_brain_states}g and Fig. \ref{Figure_6_brain_states}b), we found that the connectivity between inferior parietal lobule (node 9) and middle frontal gyrus (node 17), and the connectivity between inferior parietal lobule (node 9) and supramarginal gyrus (node 34) are increased significantly both in 2-back and 0-back working memory compared to fixation. The connectivity between angular gyrus (node 1) and supramarginal gyrus (node 34) is increased in 2-back compared to 0-back and fixation. There is reduced connectivity between the lateral occipital cortex (node 12), occipital fusiform gyrus (node 20), and  occipital pole (node 21) in 2-back and 0-back compared to fixation.

Finally, we compare 2-back place (Fig. \ref{Figure_6_brain_states}h), 0-back place (Fig. \ref{Figure_6_brain_states}k), and fixation. We found that the connectivity between lateral occipital cortex (node 12) and occipital pole (node 21), and the connectivity between occipital fusiform gyrus (node 20) and  occipital pole (node 21) are reduced in 2-back compared to 0-back and fixation.

It is clear from Fig. \ref{Figure_6_brain_states} that nodes are clustered into communities with different connectivity densities within and between communities. The mean and variance of the connectivity within and between communities are reported as block mean and variance matrices in Fig. \ref{Figure_6_brain_states}. We find that there are strong connections in communities $k$=3, 4, and 6 and weak connections in communities $k$=1, 2, and 5 for a majority of the states. The Circos map provides a different perspective on the community pattern of the brain state. We summarise the common community pattern for specific working memory load or fixation in Table \ref{table:2}.  

\begin{table*}[h!]
\centering
\begin{tabular}{llllllllllllllllll}

\multicolumn{6}{c}{\textbf{2-back}}&\multicolumn{6}{c}{\textbf{0-back}}&\multicolumn{6}{c}{\textbf{Fixation}}\\
Community&\multicolumn{4}{c}{Node number}& &Community&\multicolumn{4}{c}{Node number}& &Community&\multicolumn{4}{c}{Node number}\\
\hline
k=1& & & & & &k=1& & & & & &k=1& & & & &\\
k=2&15&30& & & &k=2& & & & & &k=2&15&31&32& &\\
k=3&16&20& & & &k=3&16&20&21& & &k=3&12&16&20&21&\\
k=4&1&9&17&34& &k=4& & & & & &k=4&1&9&25& &\\
k=5&11&14& & & &k=5& & & & & &k=5&3&11&14& &\\
k=6&8&19&35& & &k=6&19& & & & &k=6&5&8&10&19&\\
\hline
\end{tabular}
\caption{\footnotesize This table summarises the nodes commonly located in a specific community $k$ for all of the picture types in the working memory tasks.}
\label{table:2}
\end{table*}

\section*{Discussion}

We proposed a model-based method for identifying transitions and characterising brain states between two consecutive transitions. The transitions between brain states identified by the Bayesian change-point detection method exhibit appropriate lags compared to the external task demands. This indicates a significant difference between the temporal boundaries of external task demands and the transitions of latent brain states. We also estimated the community membership of brain regions that interact with each other to give rise to the brain states. Furthermore, we showed that the estimated patterns of community architectures show distinct networks for 2-back and 0-back working memory load and fixation.

We first focus on the results of the brain states inferred from the WM-tfMRI data and discuss the estimated patterns of connectivity for different blocks of working memory tasks after local inference. We find that there are distinct connectivity differences between 2-back, 0-back, and fixation. We first compare the working memory and the fixation conditions, with particular reference to the middle frontal gyrus (node 17) and inferior parietal lobule (node 9) which includes the angular gyrus (node 1) and supramarginal gyrus (node 34). The middle frontal gyrus is related to manipulation, distractor resistance, refreshing, selection for action and monitoring, and the inferior parietal lobule is related to focus recognition and long-term recollection \citep*{Nee2013}. In our results, we find that the connectivity between the middle frontal gyrus and inferior parietal lobule is increased in the working memory tasks compared to the fixation state. The connectivity between the lateral occipital cortex (node 12) and occipital fusiform cortex (node 21) is strong and stable in fixation compared to the working memory tasks, and a higher working memory load may increase the instability of this connectivity.

Regarding the difference between 2-back and 0-back working memory tasks, we focus on the angular gyrus and supramarginal gyrus. In our experimental results, we find that there is increased connectivity between the angular gyrus (node 1) and supramarginal gyrus (node 34) in 2-back compared to 0-back working memory task blocks. The angular gyrus is located in the posterior part of the inferior parietal lobule. The inferior parietal cortex, including the supramarginal gyrus and the angular gyrus, is part of a “bottom-up” attentional subsystem that mediates the automatic allocation of attention to task-relevant information \citep*{Seghier2013}. Previous work has shown that activation of the inferior parietal lobe is involved in the shifting of attention towards particular stimuli \citep*{Gottlieb2007}. The right inferior parietal lobule including angular gyrus is related to attention maintaining and salient event encoding in the environment \citep*{Singh-Curry2009}. These research findings are consistent with and justify our results. 

Next, we focus on the methodology. We introduced posterior predictive discrepancy (PPD), a novel method based on model fitness assessment combined with sliding window analysis to detect change-points in various functional brain networks and to infer the dynamics when a brain changes state. Posterior predictive assessment is a method based on Bayesian model comparison. Other Bayesian model comparison methods including Bayes factors \citep{Kass1995, West1986}, the Bayesian information criterion (BIC) \citep{Neath2012}, and Kullback–Leibler divergence \citep{Kullback1951} are also widely used in mixture modelling. One advantage of the posterior predictive assessment is that the computation for the assessment is a straightforward byproduct of the posterior sampling required in the conventional Bayesian estimation.

We defined a new criterion named cumulative discrepancy energy (CDE) to estimate locations of these change-points or transitions. The main idea underlying this novel strategy is to recognize that the goodness-of-fit between the model and observation is reduced if there is a change-point located within the current sliding window (the sample data in the window can be considered as being generated from two latent brain network architectures in this case), resulting in a significant increase in CDE. We use overlapping, rectangular, sliding windows so that all of the time points are included.

The dynamics of the brain states are not only induced by external stimuli, but also the latent mental process, such as motivation, alertness, fatigue, and momentary lapse \citep*{Taghia2018a}. Crucially, directly using the temporal boundaries (onsets and durations) associated with predefined task conditions to infer the functional networks may not be sufficiently rigorous and accurate. The boundaries of the task demand are not the timing and duration of the latent brain state. The estimated change-points in our experiments are consistent with the working memory task demands but show a delay relative to the onsets of the task blocks or the mid-points of fixation blocks. These results reflect the delay involved due to the haemodynamic response, and also the delay arising from recording the data using the fMRI scanner, between signal emission and reception. 

The results of the task fMRI data analysis show that the change-point detection algorithm is sensitive to the choice of model. We found that a less complex model (with smaller $K$) for global fitting gave fewer false positives, so it had better change-point detection performance than models with larger $K$. Selecting a suitable window size $W$ is also very important for our method. Too small a window size results in too little information being extracted from the data within the window, causing the calculated CDE to fluctuate more, making it difficult to discriminate local maxima and local minima in the CDE score time series. Too large a window size (larger than the task block length) reduces the resolution at which the change-points can be distinguished. In the working memory task fMRI data set, the length of the task block is around 34 frames and the fixation is about 20 frames. Therefore, we made the window size at most 34 frames to ensure all potential change-points can be distinguished, and at least 20 frames to ensure the effectiveness of the posterior predictive assessment. In our experiments, we used window sizes of 22, 26, 30, and 34, which were all larger than the length of the fixation block. This means it was not possible to detect the two change-points at both ends of fixation blocks, so we consider the whole fixation block as a single change-point (i.e., a buffer between two task blocks). 

The latent block model provides a flexible approach to modeling and estimating the dynamical assignment of nodes to a community. Note that the latent block model was fitted to the adjacency matrix of each individual subject in global fitting, and was fitted to the group-averaged adjacency matrix in the local fitting. Different choices of $\bm{\pi}$ can generate different connection patterns in the adjacency matrix. The likelihood is Gaussian and the connectivity is weighted, both of which facilitate treating the correlation matrix as an observation, without losing much relevant information from the time series. 

We treat both the latent label vector and block model parameters as quantities to be estimated. Changes in community memberships of the nodes are reflected in changes in the latent labels, and changes in the densities and variations in functional connectivity are reflected in changes in the model parameters. 

Empirical fMRI datasets have no ground truth regarding the locations of latent transitions of the brain states and network architectures. Although the task data experiments include the timings of stimuli, the exact change-points between discrete latent brain states are uncertain. Here, we used the multivariate Gaussian model to generate synthetic data (ground truth) to validate our proposed algorithms by comparing ground truth to the estimated change-points and latent labels. With extensive experiments using synthetic data, we demonstrated the very high accuracy of our method. The multivariate Gaussian generative model can characterize the community patterns via determining the memberships of the elements in the covariance matrix, but it is still an unrealistic benchmark. In the future, we will integrate the clustering method into the dynamic causal modelling \citep{Friston2003,Friston2019} to simulate more biologically realistic synthetic data to validate the algorithm.

There are still some limitations of the MCMC allocation sampler \citep*{Nobile2007, Wyse2012} which we use to infer the latent label vectors. When Markov chains are generated by the MCMC algorithm, the latent label vectors typically get stuck in local modes. This is in part because the Gibbs moves in the allocation sampler only update one element of the latent label vector at a time. Although the M3 move (see \textbf{Supplementary Section 7} for details on the M3 move) updates multiple elements of the latent label vector, the update is conditional on the probability ratio of a single reassignment, which results in similar problems to the Gibbs move. Improving the MCMC allocation sampler so that it can jump between different local modes, without changing the value of $K$, is a topic worth exploring. Currently, we use an MCMC sampler with a Gibbs move and an M3 move for local inference as well, keeping $K$ constant. In future work, we will extend the sampler using an absorption/ejection move, which is capable of sampling $K$ along with latent labels directly from the posterior distribution. 

The label switching phenomenon (see \textbf{Supplementary Section 2}) does not happen frequently if the chain is stuck in a local mode. However, the estimated labels in the latent label vector do switch in some experiments. To correct for label switching, we permute the labels in a post-processing stage.

In this paper, we treat the group-averaged adjacency matrix as an observation in local inference, which neglects variation between subjects \citep{Friston2016}. In the future, we propose to use hierarchical Bayesian modeling to estimate the community architecture at the group-level. In the local inference, we will model the individual adjacency matrix using the latent block model, and infer the number of communities along with the latent label vectors via an absorption/ejection strategy. At the group-level, we will model the estimated number of communities of the subjects using a Poisson-Gamma conjugate pair and model the estimated latent label vectors using a Categorical-Dirichlet pair. The posterior distribution of the number of communities will be modeled using a Gamma distribution and the posterior distribution of the latent label vector will be modelled using a Dirichlet distribution. The estimated rate of the Poisson posterior distribution and the estimated label assignment probability matrix of the Dirichlet posterior distribution will characterize the brain networks at the group-level.

The change-point detection method described in this paper can be applied to locate the relatively static brain states occurring in block designed task fMRI data. In future work, we aim to apply the method to explore the dynamic characteristics of event-related task fMRI, where applying a sliding window approach may be difficult, as the changes of the states will be the pulses. We will be also interested in applying a change-point detection algorithm to resting-state fMRI data, which is also challenging given there is no stimuli timing available and there is relatively less distinct switching of brain states.

\footnotesize

\section*{Methods}

\subsection*{Working memory task fMRI data processing}

\subsubsection*{WM-tfMRI paradigms}
The original experiment involved a version of an N-back task, used to assess working memory/cognitive control. In the working memory task, each block of tasks consisted of trials with pictures of faces, places, tools and body parts.  A specific stimulus type was presented in each block within each run. In 2-back blocks, the subjects judged whether the current stimulus is the same as the stimulus previously presented ``two back''. In 0-back blocks, the subjects were given a target cue at the beginning of each task block, and judged whether any stimulus during that block is the same as the target cue. There were 405 frames (with 0.72 s repetition time - TR) in the time course with four blocks of 2-back working memory tasks (each for 25 s), four blocks of 0-back working memory tasks (each for 25 s) and four fixation blocks (each for 15 s).
\subsubsection*{tfMRI data acquisition}
The whole brain echo-planar imaging (EPI) was acquired with a 32 channel head coil on a modified 3T Siemens Skyra with TR = 0.72 s, TE = 33.1 ms, flip angle = 52 degrees, BW = 2290 Hz/Px, in-plane FOV = $208\times 180$ mm, 72 slices with isotropic voxels of 2 mm with a multi-band acceleration factor of 8. Two runs of the tfMRI were acquired (one right to left, the other left to right).

\subsubsection*{tfMRI data preprocessing}
The tfMRI data in HCP are minimally preprocessed including gradient unwarping, motion correction, fieldmap-based EPI distortion correction, brain-boundary-based registration of EPI to structural T1-weighted scan, non-linear (FNIRT) registration into MNI152 space, and grand-mean intensity normalization. The data analysis pipeline is based on FSL (FMRIB's Software Library) \citep*{smith2004}. Further smoothing processing is conducted by Volume-based analysis and Grayordinates-based analysis, the details of which are illustrated in the corresponding sections of \citep*{Barch2013}.
\subsubsection*{GLM analysis}
The general linear model (GLM) analysis in this work includes 1st-level (individual scan run), 2nd-level (combining multiple scan runs for an individual participant) and 3rd-level (group analysis across multiple participants) analyses \citep*{Woolrich2001, Woolrich2004}. At 1st-level, fixed effects analyses are conducted to estimate the average effect size of runs within sessions, where the variation only contains the within-subject variance. At 2nd-level, we also use fixed effects analysis, averaging the two sessions within the individuals. At 3rd-level, mixed effects analyses are conducted, with the subject effect size considered to be random. The estimated mean effect size is across the population and the between subject variance is contained in the group level of GLM. We can set up different contrasts to compare the activation with respect to the memory load or stimulus type.

\subsubsection*{Time series extraction}

We created spheres of binary masks with radius 6 mm (the center of each sphere corresponded to the coordinates of locally maximum z statistics, and the voxel locations of the centers were transferred from MNI coordinates in fsleyes) and extracted the eigen time series of 35 regions of interest from the 4-D functional images. We obtained 100 sets of time series from 100 unrelated subjects using the same masks.

\subsection*{The framework of Bayesian change-point detection}
An overview of the Bayesian change-point detection framework is shown in Fig. \ref{Figure_1_schematic}a. We consider a collection of $N$ nodes $\lbrace v_{1},\cdots,v_{N} \rbrace$ representing brain regions for a single subject, and suppose that we observe a collection of $N$ time series $\textbf{Y}\in \Re^{N\times T}$ where $\textbf{Y} = (\textbf{y}_1,\textbf{y}_2,\cdots, \textbf{y}_T)$, and $T$ is the number of time points. Different background colors represent different latent network community architectures. The nodes in the networks are assumed to be clustered into communities and the different colors of the nodes represent the different community memberships. A more detailed example of changes in network architectures with 16 nodes is shown in Fig. \ref{Figure_1_schematic}b, where the community memberships are defined as a latent label vector $\textbf{z}$ and $K$ is the number of communities. A transition or change-point is defined as a time point at which the community structure changes. Correlations between time series suggest interactions between the corresponding brain regions; we therefore first process the time series to construct a sequence of graphs in which temporal correlations between time series are represented by an edge connecting the corresponding nodes. 

We apply a sliding window of width $W$ (even numbered) to the time series as shown in Fig. \ref{Figure_1_schematic}a. The sliding windows overlap and the centers of the windows are located at consecutive time points. Change-points may occur only at times $t \in \lbrace \frac{W}{2} + 1, \cdots, T - \frac{W}{2} \rbrace$ where $\frac{W}{2}$ is a margin size used to avoid computational and statistical complications. The advantage of using overlapping windows is that we can potentially detect transitions in network architecture at any time during the time course (except the margin area). For each time point $t \in \lbrace \frac{W}{2} + 1, \cdots, T - \frac{W}{2} \rbrace$, we define $\textbf{Y}_t = \lbrace\textbf{y}_{t-\frac{W}{2}}, \cdots,\textbf{y}_t,,\cdots,\textbf{y}_{t+\frac{W}{2}-1}\rbrace$ as the data in the sliding window at time $t$ and calculate a sample correlation matrix $\textbf{x}_{t}$ within this window. We interpret this correlation matrix as a weighted adjacency matrix. This means for each $t$, we obtain a sample adjacency matrix $\textbf{x}_{t}$. Subsequently, instead of time series $\textbf{Y}$, we use the sample adjacency matrix $\textbf{x}_t$ as the realized observation at time $t$. 

Fig. \ref{Figure_1_schematic}c provides a schematic illustrating the posterior predictive model fitness assessment. Specifically, we propose to use the Gaussian latent block model \citep*{Wyse2012} to quantify the likelihood of a network, and the MCMC allocation sampler (with the Gibbs move and the M3 move) \citep*{Nobile2007, Wyse2012} to infer a latent label vector $\textbf{z}$ from a collapsed posterior distribution $p(\textbf{z}|\textbf{x},K)$ derived from this model. The model parameters $\bm{\pi}$ for each block are sampled from a posterior distribution $p(\bm{\pi|\textbf{x},\textbf{z}})$, conditional on the sampled latent label vector $\textbf{z}$. The proposed model fitness procedure draws parameters (both latent label vectors and model parameters) from posterior distributions and uses them to generate a replicated adjacency matrix $\textbf{x}^{rep}$. It then calculates a disagreement index to quantify the difference between the replicated adjacency matrix $\textbf{x}^{rep}$ and realized adjacency matrix $\textbf{x}$. To evaluate model fitness, we use the parameter-dependent statistic PPDI by averaging the disagreement index. 

\subsection*{The latent block model}
The latent block model (LBM) \citep*{Wyse2012} is a random process generating networks on a fixed number of nodes $N$. The model has an integer parameter $K$, representing the number of communities. Identifying a suitable value of $K$ is a model fitting problem that will be discussed in a later section; here we assume $K$ is given. A schematic of a latent block model is shown in the brown box on the right side of Fig. \ref{Figure_1_schematic}c. A defining feature of the model is that nodes are partitioned into $K$ communities, with interactions between nodes in the same community having a different (usually higher) probability than interactions between nodes in different communities. The latent block model first assigns the $N$ nodes into the $K$ communities resulting in $K^{2}$ blocks, which are symmetric, then generates edges with a probability determined by the community memberships. The diagonal blocks represent the connectivity within the communities and the off-diagonal blocks represent the connectivity between different communities.

In this paper, we consider the edges between nodes to be weighted, so the model parameter matrix $\bm{\pi}$ consists of the means and variances that determine the connectivity in the blocks. We treat the correlation matrix as an observation, thus preserving more information from the BOLD time series than using binary edges. Given a sampled $\textbf{z}$ we can draw $\bm{\pi}$ from the posterior directly. For mathematical illustration of the latent block model, see \textbf{Supplementary section 3.1 and 3.2}.
Methods for sampling the latent vector $\textbf{z}$ will be discussed in later sections.

\subsection*{Sampling from the posterior}

The posterior predictive method we outline below involves sampling parameters from the posterior distribution. The sampled parameters are the latent label vector $\textbf{z}$ and model parameter matrix $\bm{\pi}$. There are several methods for estimating the latent labels and model parameters of a latent block model described in the literature. One method evaluated the model parameters by point estimation but considered the latent labels in $\textbf{z}$ as having a distribution \citep*{Daudin2008}, making this approach similar to an EM algorithm. Another method used point estimation for both the model parameters and latent labels \citep*{Zanghi2008}. We sample the latent label vector $\textbf{z}$ from the collapsed posterior $p(\textbf{z}\vert \textbf{x},K)$ (see detailed derivation of $p(\textbf{z}\vert \textbf{x},K)$ in \textbf{Supplementary Section 3.3}). We use the Markov chain Monte Carlo (MCMC) \citep*{Hastings1970} method to sample the latent label vector from the posterior using Gibbs moves and M3 moves \citep*{Nobile2007} for updating $\textbf{z}$. The details of the MCMC allocation sampler and the computational complexity are illustrated in \textbf{Supplementary Section 3.4}. After sampling the latent label vector $\textbf{z}$, we then separately sample $\bm{\pi}$ from the density $p(\bm{\pi}\vert\textbf{x},\textbf{z})$ (See \textbf{Supplementary section 3.2} for the details).

\subsection*{Model fitting}

\subsubsection*{Global fitting}
\textit{Global fitting} uses a model with a constant number of communities $K$ to fit consecutive individual adjacency matrices within a sliding window, for all time frames. For global fitting, we consider $K$ in our latent block model to be fixed over the time course. We detect the change-points based on Bayesian model comparison using posterior predictive discrepancy, which does not determine whether the model is `true' or not, but rather quantifies the preference for the model given the data. One can imagine the model as a moving ruler under the sliding window, and the observation at each time step as the object to be measured. The discrepancy increases significantly if there is a change-point located within the window. Although $K$ is constant in global fitting, different values of $K$ can be used if we select different models. The evaluation of $K$ can be considered as a Bayesian model comparison problem. We repeat the inference with different values of $K$ and compare the change-point detection performance to identify an appropriate value for $K$.
\subsubsection*{Local fitting}
\textit{Local fitting} involves first selecting a model (i.e., choosing a value of $K$) that best fits the group averaged adjacency matrix for a discrete brain state. Subsequently, the data between change-points is used to estimate the community membership that constitutes that discrete brain state. We treat $K$ as constant for this \textit{local inference}. The number of communities $K$ can potentially be inferred using the absorption/ejection move \citep*{Nobile2007} in the allocation sampler, an innovation that will be explored in future research.
 
\subsection*{Posterior predictive discrepancy}
Given inferred values of $\textbf{z}$ and $\bm{\pi}$ under the model $K$, one can draw a replicated adjacency matrix $\textbf{x}^{rep}$ from the predictive distribution $P(\textbf{x}^{rep}\vert\textbf{z},\bm{\pi},K)$ as shown in Fig. \ref{Figure_1_schematic}c. Note that the realized adjacency matrix (i.e., an observation) and the replicated adjacency matrix are conditionally independent,
\begin{equation}
P( \textbf{x},\textbf{x}^{rep}\vert\textbf{z},\bm{\pi},K)=P(\textbf{x}^{rep}\vert\textbf{z},\bm{\pi},K)P(\textbf{x}\vert\textbf{z},\bm{\pi},K).
\end{equation} 
Multiplying both sides of this equality by $P(\textbf{z},\bm{\pi}\vert\textbf{x},K) / P(\textbf{x}\vert\textbf{z},\bm{\pi},K)$ gives
\begin{equation}
P(\textbf{x}^{rep},\textbf{z},\bm{\pi}\vert \textbf{x},K)=P(\textbf{x}^{rep}\vert\textbf{z},\bm{\pi},K)P(\textbf{z},\bm{\pi}\vert\textbf{x},K).
\end{equation}

Here we use a replicated adjacency matrix in the context of posterior predictive assessment \citep*{Gelman1996} to evaluate the fitness of a posited latent block model to a realized adjacency matrix. We generate a replicated adjacency matrix by first drawing samples ($\textbf{z}$, $\bm{\pi}$) from the joint posterior $P(\textbf{z},\bm{\pi}\vert\textbf{x},K)$. Specifically, we sample the latent label vector $\textbf{z}$ from $p(\textbf{z}\vert \textbf{x},K)$ and model parameter $\bm{\pi}$ from $p(\bm{\pi}\vert\textbf{x},\textbf{z})$ and then draw a replicated adjacency matrix from $P(\textbf{x}^{rep}\vert\textbf{z},\bm{\pi},K)$. We  compute a discrepancy function to assess the averaged difference between the replicated adjacency matrix $\textbf{x}^{rep}$ and the realized adjacency matrix $\textbf{x}$, as a measure of model fitness. 

In \citep*{Gelman1996}, the $\chi^{2}$ function is used as the discrepancy measure, where the observation is considered as a vector. However, in the latent block model, the observation is a weighted adjacency matrix and the sizes of the sub-matrices can vary. In this paper, we propose a new discrepancy index to compare adjacency matrices $\textbf{x}^{rep}$ and $\textbf{x}$. We define a disagreement index to evaluate the difference between the realized adjacency matrix and the replicated adjacency matrix. This disagreement index is denoted $\gamma(\textbf{x}^{rep};\textbf{x})$ and can be considered as a parameter-dependent statistic.    
In mathematical notation, the disagreement index $\gamma$ is defined as
\begin{equation}
\gamma(\textbf{x}^{rep};\textbf{x})=\frac{\sum_{i=1,j=1}^{N} {\vert\textbf{x}_{ij}-\textbf{x}_{ij}^{rep}\vert}}{N^{2}},
\end{equation}
For the evaluation of model fitness, we generate $S$ replicated adjacency matrices and define the posterior predictive discrepancy index (PPDI) $\overline{\gamma}$ as follows. 
\begin{equation}
\overline{\gamma}=\frac{\sum_{i=1}^{S} \gamma(\textbf{x}^{rep^{i}};\textbf{x})}{S}.
\end{equation}

The computational cost of the posterior predictive discrepancy procedure in our method depends mainly on two aspects. The first is the iterated Gibbs and M3 moves used to update the latent variable vectors. The computational cost of these moves has been discussed in previous sections. The second aspect is the number of replications $S$ needed for the predictive process. Posterior predictive assessment is not sensitive to the replication number $S$, but $S$ linearly impacts the computational cost, that is, the computational complexity of model fitness assessment is $O(S)$. There is a natural trade-off between increasing the replication number and reducing the computational speed. 

\subsection*{Cumulative discrepancy energy}

Our proposed strategy to detect network community change-points is to assess the fitness of a latent block model by computing the posterior predictive discrepancy index (PPDI) $\overline{\gamma}_t$ for each $t \in \lbrace \frac{W}{2} + 1, \cdots, T - \frac{W}{2} \rbrace$. The key insight here is that the fitness of the model is relatively worse when there is a change-point within the window used to compute $\textbf{x}_t$. If there is a change-point within the window, the data observed in the left and right segments are generated by different network architectures, resulting in poor model fit and a correspondingly high posterior predictive discrepancy index. 

In practice, we find that the PPDI fluctuates severely. To identify the most plausible position of a change-point, we use another window with window size $W_{s}$ to accumulate the PPDI time series. We obtain the cumulative discrepancy energy (CDE) $E_{t}$, given by
\begin{equation}
E_{t}=\sum_{i=t-\frac{W_{s}}{2}}^{t+\frac{W_{s}}{2}-1}\overline{\gamma}_i.
\end{equation}
We take the locations of change-points to be the local maxima of the cumulative discrepancy energy, where those maxima rise sufficiently high above the surrounding sequence. The change-point detection algorithm is summarized in \textbf{Supplementary Section 8}.

Note that the posterior predictive discrepancy index and cumulative discrepancy energy for change-point detection are calculated under the conditions of global fitting. For group analysis, we average CDEs across subjects to obtain the Group CDE. After discarding false positives, the change-points are taken to be the local maxima and the discrete states are inferred at the local minima. 

\subsection*{Local inference}
We estimate the community structure of brain states via local inference. For local inference, we first calculate the group averaged adjacency matrix of 100 subjects using the data between two estimated change-points and treat this as an observation of each discrete brain state, then we use local fitting to select a value $K$ using the latent block model for Bayesian estimation of community structure for each brain state.

\subsection*{Code availability}

The code for GLM analysis (Shell script), Bayesian change-point detection (MATLAB), and brain network visualization (MATLAB, Perl) is available at: \href{https://github.com/LingbinBian/BCPD_v1.0}{https://github.com/LingbinBian/BCPD1.0}.   

%

\newpage
\onecolumn

\large

{\centering

\Huge \textbf{Supplementary information}}
\\---------------------------------------------------------------------------------------------------------------------------------------\\
\\
\\
\captionsetup[figure]{labelfont={bf},name={Supplementary Figure},labelsep=period}
\section{Generative model, synthetic data and parameter settings}
To validate our Bayesian change-point detection algorithm, we use the multivariate Gaussian generative model to simulate the synthetic data. Specifically, we generate $D$ segments of Gaussian time series from $D$ different network architectures. The synthetic data contains the ground truth of $D-1$ change-points over the time course. The positions of the true change-points are denoted as a row vector $\textbf{p}=[p_{1},\cdots,p_{D-1}]$. Within each of $D$ segments, we suppose nodes are assigned to $K^{true}$ communities, the value of which differs in different segments. The true number of communities in the segments can be denoted as a vector $\textbf{K}^{true}=[K_{1}^{true},\cdots,K_{D}^{true}]$. We set the label vectors that determine the form of the covariance matrices in the generative model to be $\lbrace \textbf{z}_1, \textbf{z}_2,\cdots,\textbf{z}_D \rbrace$. These label vectors are generated using the Dirichlet-Categorical conjugate pair. The component weights $\lbrace \textbf{r}_1, \textbf{r}_2,\cdots,\textbf{r}_D \rbrace$ are first drawn from a uniform distribution on the $\textbf{K}^{true}$ simplex and then nodes are assigned to the communities by drawing from the corresponding Categorical distributions.  Time series data in $\Re^{N}$ are then simulated from
\begin{equation}
Y=f(\textbf{z},a,b)+\bm{\epsilon}
\end{equation}
for $t = 1, \cdots, T$ by drawing $f(\textbf{z},a,b)\sim\mathcal{N}(\textbf{0},\bm{\Sigma}(\textbf{z},a,b))$, with 
\begin{equation}
\Sigma_{ij}=\left\{
             \begin{array}{lr}
             1, \ \ \ \ \mbox{if}\ i\ = j\ \\
             a,\ \ \  \ \mbox{if}\ i\ \neq j\ \mbox{and}\ z_{i}=z_{j} \\    
             b,\ \ \  \ \mbox{if}\ i\ \neq j\ \mbox{and}\ z_{i}\neq z_{j}\\
             \end{array}
\right.
\end{equation}
where $a\sim U(0.8,1)$ and $b\sim U(0,0.2)$ are uniformly distributed, and $\bm{\epsilon}\sim \mathcal{N}(\textbf{0},\sigma^{2}\textbf{I})$ is the additive Gaussian noise. The resulting covariance matrices for $D$ segments are denoted as $\lbrace \bm{\Sigma}_{1}, \bm{\Sigma}_{2},\cdots,\bm{\Sigma}_{D} \rbrace$. The simulated data $\textbf{Y}\in \Re^{N\times T}$ can be separated into $D$ segments which are  $\lbrace \textbf{Y}_1, \textbf{Y}_2,\cdots,\textbf{Y}_D \rbrace$. 

For validation, we first generate 100 instances (as virtual subjects) of synthetic multivariate time series for a network with $N = 35$ nodes and $T=180$ time points to imitate the scenario of real data. We set the true change-points at $\lbrace 20, 50, 80, 100, 130, 160 \rbrace$ and the numbers of communities in the segments to be $\lbrace 3,4,5,3,5,4,3\rbrace$. Here we define the signal-to-noise ratio (SNR) as $\frac{\Sigma_{ii}}{\sigma^{2}}$, and set different values of $\sigma$ to control SNR ($\sigma = 0.3162$ for SNR = 10dB, $\sigma = 0.5623$ for SNR = 5dB, $\sigma = 1$ for SNR = 0dB, and $\sigma = 1.7783$ for SNR = -5dB). For global fitting, the posterior prediction replication number is set as $S=50$ for all of our experiments. For local inference, we draw $S_{s}=200$ samples from the posterior densities for both latent label vectors and model parameters. We set the prior to be $\mbox{NIG}(\xi,\kappa^2\sigma_{kl}^{2},\nu/2,\rho/2)$ with $\xi=0$, $\kappa^{2}=1$, $\nu=3$ and $\rho=0.02$, which is non-informative.

\section{Label switching}
For the latent block model, we set $\alpha_{k}=1$ with $\lbrace k=1,\cdots,K\rbrace$, and constant values of $\xi$, $\kappa^{2}$, $\nu$ and $\rho$ for all of the blocks $kl$, so the prior is symmetric with respect to permutations of community labels. Permutations of community labels do not change the likelihood, which means the distributions with respect to blocks are not identifiable. 
Therefore, the posterior is also invariant to permutations of community labels. 
In the Markov chain, the labels of the latent label vector switch occasionally: this effect is known as the label switching phenomenon \citep*{Stephens2000, Nobile2007, Wyse2012}. For global fitting, label switching does not affect the results of posterior predictive discrepancy. However, for local inference, we need to assign the labels to the communities unequivocally to estimate the memberships of the nodes. 

We define a distance indicating the difference of coordinates between two latent label vectors $\textbf{z}$ and $\textbf{z}'$, 
\begin{equation}
D(\textbf{z},\textbf{z}')=\sum_{i=1}^{N}I(z_{i}\neq z'_{i}),
\end{equation}
where $I$ is the indicator function.
We define
\begin{equation}
\bm{\sigma}=\lbrace\sigma(1),\cdots,\sigma(k),\cdots,\sigma(K)\rbrace
\end{equation}
as a permutation of a labelling $\lbrace 1,\cdots,k,\cdots, K\rbrace$.
Let $\textbf{Q}=\lbrace\textbf{z}^{j}(\bm{\sigma}^{j}), j=1,\cdots,J\rbrace$ be a collection of latent label vectors with respect to a sequence of permutations $\lbrace\bm{\sigma}^j, j=1,\cdots,J\rbrace$. We want to minimize the sum of all distances between the vectors\begin{equation}
\sum_{j=1}^{J-1}\sum_{l=j+1}^{J} D(\textbf{z}^{j}(\bm{\sigma}^{j}),\textbf{z}^{l}(\bm{\sigma}^{l})).
\end{equation}
The solution of this  minimization can be considered as a sequential optimization problem of the square assignment. For each vector $\textbf{z}^{j}$, if the vectors that have already been processed (relabelled) up to $j-1$ are $\lbrace\textbf{z}^{t},t=1,\cdots,j-1\rbrace$, we define the element of a cost matrix
\begin{equation}
C(k_{1},k_{2})=\sum_{t=1}^{j-1}\sum_{i=1}^{N} D(z_{i}^{t}\neq k_{1},z_{i}^{j}= k_{2}).
\end{equation}
We use the square assignment algorithm \citep*{Carpaneto1980} returning a permutation $\bm{\sigma}^{j}$ which minimizes the total cost $\sum_{k=1}^{K}C(k,\sigma(k))$ for each $\textbf{z}^{j}$. Finally, we permute the labels in the vector $\textbf{z}^{j}$ according to $\bm{\sigma}^{j}$.

\section{Bayesian Modelling for functional connectivity}

\subsection{Clustering with latent block model}
Mathematically, we denote the community memberships (also called the latent labels) of the nodes as a vector $\textbf{z} = (z_1, \ldots, z_N)$ such that $z_{i}\in\lbrace 1,\cdots,K \rbrace$ denotes the community containing node $i$. Each $z_{i}$ independently follows the categorical (one-trial multinomial) distribution: 
\begin{equation}
z_{i}\sim \mbox{ Categorical}(1; \textbf{r}=\lbrace r_{1},\cdots,r_{K}\rbrace),
\end{equation}
where $r_{k}$ is the probability of a node being assigned to community $k$ and $\sum_{k=1}^{K} r_{k}=1$. The categorical probability can be expressed using the indicator function $I_{k}(z_{i})$ as
\begin{equation}
p(z_{i}\vert\textbf{r},K)=\prod_{k=1}^{K}r_{k}^{I_{k}(z_{i})}, \mbox{where\ }  I_{k}(z_{i})=
\begin{cases}
1, \ \mbox{if}\ z_{i}=k\\
0, \ \mbox{if}\ z_{i}\neq k\\
\end{cases}.
\end{equation}
This implies that the $N$ dimensional vector $\textbf{z}$ is generated with probability
\begin{equation}
p(\textbf{z}\vert\textbf{r},K)=\prod_{k=1}^{K}r_{k}^{m_k(\textbf{z})},
\end{equation}
where $m_{k}(\textbf{z})=\sum_{i=1}^{N}I_{k}(z_{i})$. The latent allocation parameter vector $\textbf{r}=(r_{1},\cdots,r_{K})$ is assumed to have a $K$-dimensional Dirichlet prior with density 
 
\begin{equation}
p(\textbf{r}\vert K)=N(\bm{\alpha})\prod_{k=1}^{K}r_{k}^{\alpha_{k}-1}, 
\end{equation}
where the normalization factor is $N(\bm{\alpha})=\frac{\Gamma(\sum_{k=1}^{K}\alpha_{k})}{\prod_{k=1}^{K}\Gamma(\alpha_{k})}$. In this work we suppose $\alpha_k = 1$ for $k = 1, \ldots, K$, so that the prior for $\textbf{r}$ is uniform on the $K$-simplex. 
Edges between nodes are represented using an adjacency matrix $\textbf{x}\in \Re^{N\times N}$. We define a block $\textbf{x}_{kl}$ comprised of weighted edges connecting the nodes in community $k$ to the nodes in community $l$. The likelihood of the latent block model can be expressed as
\begin{equation}
p(\textbf{x}\vert \bm{\pi},\textbf{z},K)=\prod_{k.l}p(\textbf{x}_{kl}\vert \pi_{kl},\textbf{z},K),
\end{equation}
and the likelihood in specific blocks can be expanded as
\begin{equation}
p(\textbf{x}_{kl}\vert \pi_{kl},\textbf{z},K)=\prod_{\lbrace i\vert z_{i}=k \rbrace}\prod_{\lbrace j\vert z_{j}=l \rbrace} p(x_{ij}\vert \pi_{kl},\textbf{z},K),
\end{equation}
where $\bm{\pi}=\lbrace \pi_{kl} \rbrace$ is a $K\times K$ model parameter matrix. 

\subsection{The latent block model with weighted edges}

The block model parameter in block $kl$ is $\pi_{kl}=(\mu_{kl},\sigma_{kl}^{2})$ and each $x_{ij}$ in the block $kl$ follows a Gaussian distribution conditional on $\textbf{z}$ under the model $K$, that is 
\[
x_{ij}\vert\pi_{kl}, \textbf{z},K \sim \mathcal{N}(\mu_{kl},\sigma_{kl}^{2}).
\]
The parameter vectors $\pi_{kl}=(\mu_{kl},\sigma_{kl}^{2})$ are assumed to independently follow the conjugate Normal-Inverse-Gamma (NIG) prior $\pi_{kl}\sim \mbox{NIG}(\xi,\kappa^2\sigma_{kl}^{2},\nu/2,\rho/2)$. That is, $\mu_{kl}\sim\mathcal{N}(\xi,\kappa^2\sigma_{kl}^{2})$ and $\sigma_{kl}^{2}\sim \mbox{IG}(\nu/2,\rho/2)$. The density of the Inverse-Gamma distribution $\mbox{IG}(\alpha,\beta)$ has the general formula
$
p(x)=\frac{\beta^{\alpha}}{\Gamma(\alpha)}x^{-(\alpha+1)}e^{(\frac{-\beta}{x})}
$, where $\alpha$ and $\beta$ are hyper-parameters.

We define $s_{kl}(\textbf{x})$ to be the sum of the edge weights in the block $kl$ and $q_{kl}(\textbf{x})$ to be the sum of squares as follows:
\begin{equation}
s_{kl}(\textbf{x})=\sum_{i:z_{i}=k}\sum_{j:z_{j}=l}x_{ij},
\end{equation}
and
\begin{equation}
q_{kl}(\textbf{x})=\sum_{i:z_{i}=k}\sum_{j:z_{j}=l}x_{ij}^{2}.
\end{equation}

We also define $w_{kl}(\textbf{z})=m_{k}(\textbf{z})m_{l}(\textbf{z})$ to be the number of elements in the block, where $m_k$ and $m_l$ are the numbers of nodes in community $k$ and $l$ respectively. The prior and the likelihood in the above expression is the NIG-Gaussian conjugate pair. With this conjugate pair, we can calculate the posterior distribution for each model block, which is also a Normal-Inverse-Gamma distribution 
$\mu_{kl}\sim\mathcal{N}(\xi_n,\kappa_n^2\sigma_{kl}^{2})$ and $\sigma_{kl}^{2}\sim \mbox{IG}(\nu_n/2,\rho_n/2)$, where
\begin{equation}
\nu_n=\nu+w_{kl},
\end{equation}
\begin{equation}
\kappa_n^2=\frac{\kappa^2}{1+w_{kl}\kappa^2},
\end{equation}
\begin{equation}
\xi_n=\frac{\xi+s_{kl}\kappa^2}{1+w_{kl}\kappa^2},
\end{equation}
\begin{equation}
\rho_n=\frac{\xi^{2}}{\kappa^{2}}+q_{kl}+\rho-\frac{(\xi+s_{kl}\kappa^{2})^2}{1/\kappa^{2}+w_{kl}}.
\end{equation}
Details of the derivation of this  $\mbox{NIG}(\xi_n,\kappa_n^2\sigma_{kl}^{2},\nu_n/2,\rho_n/2)$ distribution are provided in \textbf{Supplementary Section 4}. The posterior density of the whole model is a product of such terms for all blocks, as follows.  
\begin{equation}
p(\bm{\pi}\vert\textbf{x},\textbf{z})=\prod_{k,l}p(\pi_{kl}\vert\textbf{x}_{kl},\textbf{z}).
\end{equation} 
Given a sampled $\textbf{z}$ we can draw $\bm{\pi}$ from the above posterior directly. 
Methods for sampling the latent vector $\textbf{z}$ will be discussed later in the paper.

\subsection{The collapsed posterior of latent label vector}
In this model, a change-point corresponds to a change in community architecture i.e., a change in the latent label vector $\textbf{z}$ and the parameter matrix $\bm{\pi}$. For the sake of computational efficiency, it is convenient to construct the collapsed posterior distribution $p(\textbf{z}\vert \textbf{x},K)$. We can obtain the collapsed posterior by integrating out the nuisance parameters \citep*{MacDaid2012, Wyse2012}. In this section, we discuss the details of collapsing the latent block model when the edge weights are continuously valued. 

Given $K$, 
the joint density of $\textbf{x}$, $\bm{\pi}$, $\textbf{z}$, and $\textbf{r}$ is 
\begin{equation}
p(\textbf{x},\bm{\pi},\textbf{z},\textbf{r}\vert K)=p(\textbf{z},\textbf{r}\vert K)p(\textbf{x},\bm{\pi}\vert \textbf{z}).
\end{equation}
The parameters $\textbf{r}$ and $\bm{\pi}$ can be integrated out (collapsed) to obtain the marginal density $p(\textbf{x},\textbf{z}\vert K)$. 
\begin{equation}
p(\textbf{z},\textbf{x}\vert K)=\int p(\textbf{z},\textbf{r}\vert K)d\textbf{r}\int p(\textbf{x},\bm{\pi}\vert \textbf{z})d\bm{\pi},
\end{equation}
so that the posterior for the block-wise model can be expressed as
\begin{equation}
p(\textbf{z}\vert \textbf{x},K)\propto p(\textbf{z}, \textbf{x}\vert K)= \int p(\textbf{z},\textbf{r}\vert K)d\textbf{r}  \prod_{k,l}\int p(\textbf{x}_{kl},\pi_{kl}\vert \textbf{z})d\pi_{kl}.
\end{equation}
The first integral $p(\textbf{z}\vert K)=\int p(\textbf{z},\textbf{r}\vert K)d\textbf{r}$, where the integral is over the $K$-simplex, can be evaluated as follows:
\begin{eqnarray}
\int p(\textbf{z},\textbf{r}\vert K)d\textbf{r}&=&\frac{\Gamma(\sum_{k=1}^{K}\alpha_{k})}{\Gamma(\sum_{k=1}^{K}(\alpha_{k}+m_{k}(\textbf{z}))}\\
&&\times\prod_{k=1}^{K}\frac{\Gamma(\alpha_{k}+m_{k}(\textbf{z}))}{\Gamma(\alpha_{k})}.
\end{eqnarray}
The details of this derivation are in \textbf{Supplementary Section 5} below. The integral of the form $\int p(\textbf{x}_{kl},\pi_{kl}\vert \textbf{z})d\pi_{kl}$ can be evaluated as
\begin{eqnarray}
\int p(\textbf{x}_{kl},\pi_{kl}\vert \textbf{z})d\pi_{kl}&=&\frac{\rho^{\nu/2}\Gamma\lbrace(w_{kl}+\nu)/2\rbrace}{\pi^{w_{kl}/2}\Gamma(\nu/2)(w_{kl}\kappa^2+1)^{1/2}}\\
&&\times(-\frac{\kappa^2(s_{kl}+\xi/\kappa^2)^2}{w_{kl}\kappa^2+1}+\frac{\xi^2}{\kappa^2}\\
&&+q_{kl}+\rho)^{-(w_{kl}+\nu)/2}
\end{eqnarray}
The derivation is in \textbf{Supplementary Section 6}. 

\subsection{Sampling from the collapsed posterior}
We use a Markov chain Monte Carlo (MCMC) method to sample the latent label vector from the posterior with proposal moves $p(\textbf{z}\rightarrow\textbf{z}^{\ast})$ similar to those of the allocation sampler \citep*{Nobile2007} to update $\textbf{z}$. In the Metropolis-Hastings algorithm \citep*{Hastings1970}, a candidate latent label vector $\textbf{z}^{\ast}$ is accepted with probability $\min\lbrace 1,r \rbrace$, where
\begin{equation}
r=\frac{p(K,\textbf{z}^{\ast},\textbf{x})p(\textbf{z}^{\ast}\rightarrow\textbf{z})}{p(K,\textbf{z},\textbf{x})p(\textbf{z}\rightarrow\textbf{z}^{\ast})}.
\end{equation} 
In each iteration of the sampler, we perform either a Gibbs move or an M3 move, with equal probability (0.5) of each. Each Gibbs move updates the latent label vector $\textbf{z}$ by drawing from the collapsed posterior $p(\textbf{z}\vert \textbf{x},K)$. At each iteration, one entry $z_{i}$ is randomly selected and updated by drawing from
\begin{equation}
p(z_{i}^{\ast}\vert z_{-i},\textbf{x},K)=\frac{1}{C} p(z_{1},\cdots,z_{i-1},z_{i}^{\ast}=k,z_{i+1},\cdots,z_{n}\vert\textbf{x}),  
\end{equation}
where $k\in \lbrace 1,\cdots,K \rbrace$, $z_{-i}$ represents the elements in $\textbf{z}$ apart from $z_{i}$ and the normalization term
\begin{equation}
C=p(z_{-i}\vert\textbf{x},K)=\sum_{k=1}^{K}p(z_{1},\cdots,z_{i-1},z_{i}^{\ast}=k,z_{i+1},\cdots,z_{n}\vert\textbf{x}).
\end{equation}
For a Gibbs move within a Metropolis-Hastings sampler, the ratio $r$ always equals one. The computational complexity of a Gibbs move depends on the cost of calculating the probability of the reassignment of a specific entry. Each probability takes $O(K^2+N^2)$ time to calculate. There are $K$ possible reassignments so that each Gibbs move takes $O(K^3+KN^2)$ time.

The details of the M3 move are provided in \textbf{Supplementary Section 7}. The computational complexity of the M3 move depends on the cost of calculating the ratio of posterior density and proposal density. The time cost of calculating this ratio is $O(K^2+N^2)$, and calculating the proposal ratio takes $O(N+L^2)$ time, so the M3 move takes $O(K^2+N^2+L^2)$ time.

\section{The likelihood and posterior of the latent block model with weighted edges}
\textbf{Likelihood}: The likelihood of the block $kl$ with weighted edges is
\begin{eqnarray}
p(\textbf{x}_{kl}\vert \pi_{kl},\textbf{z},K)
&=&\prod_{\lbrace i\vert z_{i}=k \rbrace}\prod_{\lbrace j\vert z_{j}=l \rbrace} p(x_{ij}\vert \mu_{kl},\sigma_{kl}^{2},\textbf{z},K) \nonumber  \\
&=&(2\pi\sigma_{kl}^{2})^{-w_{kl}/2}\mbox{exp}\lbrace -\frac{1}{2\sigma_{kl}^2}\sum_{i:z_{i}=k}\sum_{j:z_{j}=l}(x_{ij}-\mu_{kl})^2\rbrace \nonumber \\
&=&(2\pi\sigma_{kl}^{2})^{-w_{kl}/2} \nonumber \\
&&\times\mbox{exp}\lbrace -\frac{1}{2\sigma_{kl}^2}(\sum_{i:z_{i}=k}\sum_{j:z_{j}=l}x_{ij}^2-2\sum_{i:z_{i}=k}\sum_{j:z_{j}=l}x_{ij}\mu_{kl}\nonumber \\
& &+\sum_{i:z_{i}=k}\sum_{j:z_{j}=l}\mu_{kl}^2)\rbrace\nonumber \\
&=&(2\pi\sigma_{kl}^{2})^{-w_{kl}/2}\mbox{exp}\lbrace -\frac{1}{2\sigma_{kl}^2}(q_{kl}-2\mu_{kl}s_{kl}+w_{kl}\mu_{kl}^2)\rbrace,
\end{eqnarray}
where $w_{kl}$ is the number of elements in block $kl$, $s_{kl}$ is the sum of the weights and $q_{kl}$ is the sum of squares of the weights in the block $kl$.

\textbf{Posterior}: We derive the posterior of the model parameter $\pi_{kl}$ with prior $\mu_{kl}\sim\mathcal{N}(\xi,\kappa^2\sigma_{kl}^{2})$ and $\sigma_{kl}^{2}\sim \mbox{IG}(\nu/2,\rho/2)$ as follows.
\begin{eqnarray}
p(\pi_{kl}\vert\textbf{x}_{kl},\textbf{z},K)&\propto& p(\pi_{kl})p(\textbf{x}_{kl}\vert\pi_{kl}, \textbf{z},K)\nonumber \\
&=& p(\mu_{kl})p(\sigma_{kl}^2)\prod_{\lbrace i\vert z_{i}=k \rbrace}\prod_{\lbrace j\vert z_{j}=l \rbrace} p(x_{ij}\vert \mu_{kl},\sigma_{kl}^{2},\textbf{z},K)\nonumber \\
&=&(2\pi\kappa^2\sigma_{kl}^2)^{-1/2}\mbox{exp}\lbrace-\frac{1}{2\kappa^2\sigma_{kl}^2}(\mu_{kl}-\xi)^2 \rbrace\nonumber \\
&&\times\frac{(\rho/2)^{\nu/2}}{\Gamma(\nu/2)}\sigma_{kl}^{-2(\nu/2+1)}\mbox{exp}\lbrace-\rho/2\sigma_{kl}^2\rbrace\nonumber \\
&&\times(2\pi\sigma_{kl}^{2})^{-w_{kl}/2}\mbox{exp}\lbrace -\frac{1}{2\sigma_{kl}^2}(q_{kl}-2\mu_{kl}s_{kl}+w_{kl}\mu_{kl}^2)\rbrace\nonumber \\
&=&\frac{(\rho/2)^{\nu/2}}{\Gamma(\nu/2)}(2\pi\kappa^2)^{-1/2}(2\pi)^{-w_{kl}/2}\sigma_{kl}^{-1}\sigma_{kl}^{-\nu-2-w_{kl}}\nonumber \\
&&\times\mbox{exp}\lbrace -\frac{1}{2\sigma_{kl}^2}[(\frac{1}{\kappa^2}+w_{kl})\mu_{kl}^2-2(\frac{1}{\kappa^2}\xi+s_{kl})\mu_{kl}\nonumber \\
&&+\frac{1}{\kappa^2}\xi^2+q_{kl}+\rho]\rbrace
\end{eqnarray}
The posterior of the Gaussian model is also a Normal-Inverse-Gamma distribution which can be denoted as
$\mu_{kl}\sim\mathcal{N}(\xi_n,\kappa_n^2\sigma_{kl}^{2})$ and $\sigma_{kl}^{2}\sim \mbox{IG}(\nu_n/2,\rho_n/2)$. The posterior density can be expressed as
\begin{eqnarray}
p(\pi_{kl}\vert\textbf{x}_{kl},\textbf{z},K)&=&(2\pi\kappa_n^2\sigma_{kl}^2)^{-1/2}\mbox{exp}\lbrace-\frac{1}{2\kappa_n^2\sigma_{kl}^2}(\mu_{kl}-\xi_n)^2 \rbrace\nonumber \\
&&\times\frac{(\rho_n/2)^{\nu_n/2}}{\Gamma(\nu_n/2)}\sigma_{kl}^{-2(\nu_n/2+1)}\mbox{exp}\lbrace-\rho_n/2\sigma_{kl}^2\rbrace\nonumber \\
&=&\frac{(\rho_n/2)^{\nu_n/2}}{\Gamma(\nu_n/2)}(2\pi\kappa_n^2)^{-1/2}\sigma_{kl}^{-1}\sigma_{kl}^{-\nu_n-2}\nonumber \\
&&\times \mbox{exp}\lbrace -\frac{1}{2\sigma_{kl}^2}(\frac{1}{\kappa_n^2}\mu_{kl}^2-\frac{2\xi_n}{\kappa_n^2}\mu_{kl}+\frac{\xi_n^2}{\kappa_n^2}+\rho_n)\rbrace.
\end{eqnarray}
Comparing the terms and coefficients with respect to $\mu_{kl}^2$, $\mu_{kl}$ and $\sigma_{kl}^2$, 
\begin{equation}
-\nu_n-2=-\nu-2-w_{kl},
\end{equation}
\begin{equation}
\frac{1}{\kappa_n^2}=\frac{1}{\kappa^2}+w_{kl},
\end{equation}
\begin{equation}
\frac{2\xi_n}{\kappa_n^2}=2(\frac{1}{\kappa^2}\xi+s_{kl}),
\end{equation}
\begin{equation}
\frac{\xi_n^2}{\kappa_n^2}+\rho_n=\frac{1}{\kappa^2}\xi^2+q_{kl}+\rho.
\end{equation}
In summary, the parameters of the posterior density are given by
\begin{equation}
\nu_n=\nu+w_{kl},
\end{equation}
\begin{equation}
\kappa_n^2=\frac{\kappa^2}{1+w_{kl}\kappa^2},
\end{equation}
\begin{equation}
\xi_n=\frac{\xi+s_{kl}\kappa^2}{1+w_{kl}\kappa^2},
\end{equation}
\begin{equation}
\rho_n=\frac{\xi^{2}}{\kappa^{2}}+q_{kl}+\rho-\frac{(\xi+s_{kl}\kappa^{2})^2}{1/\kappa^{2}+w_{kl}}.
\end{equation}
We can directly sample $\pi_{kl}$ from $\mbox{NIG}(\xi_n,\kappa_n^2\sigma_{kl}^{2},\nu_n/2,\rho_n/2)$.\\

\section{Collapse $\textbf{r}$ in latent block model}
We show the calculation of $p(\textbf{z}\vert K)=\int p(\textbf{z},\textbf{r}\vert K)d\textbf{r}$. Given the $K$-dimensional Dirichlet prior with density $p(\textbf{r}\vert K)=N(\bm{\alpha})\prod_{k=1}^{K}r_{k}^{\alpha_{k}-1}$, where $\bm{\alpha}=\lbrace\alpha_{1},\cdots,\alpha_{K}\rbrace$, $N(\bm{\alpha})=\frac{\Gamma(\sum_{k=1}^{K}\alpha_{k})}{\prod_{k=1}^{K}\Gamma(\alpha_{k})}$; and the likelihood $p(\textbf{z}\vert\textbf{r},K)=\prod_{k=1}^{K}r_{k}^{m_k(\textbf{z})}$, we can collapse $\textbf{r}$ as follows:
\begin{eqnarray}
p(\textbf{z}\vert K)&=&\int p(\textbf{z},\textbf{r}\vert K)d\textbf{r}\nonumber \\
&=&\int p(\textbf{r}\vert K)p(\textbf{z}\vert\textbf{r}, K)d\textbf{r}\nonumber \\
&=&\int \frac{\Gamma(\sum_{k=1}^{K}\alpha_{k})}{\prod_{k=1}^{K}\Gamma(\alpha_{k})}\prod_{k=1}^{K}r_{k}^{\alpha_{k}-1} \prod_{k=1}^{K}r_{k}^{m_k}d\textbf{r}\nonumber \\
&=&\frac{\Gamma(\sum_{k=1}^{K}\alpha_{k})}{\prod_{k=1}^{K}\Gamma(\alpha_{k})}\frac{\prod_{k=1}^{K}\Gamma(\alpha_{k}+m_{k})}{\Gamma(\sum_{k=1}^{K}(\alpha_{k}+m_{k}))}\nonumber \\
&&\times \int \frac{\Gamma(\sum_{k=1}^{K}(\alpha_{k}+m_{k}))}{\prod_{k=1}^{K}\Gamma(\alpha_{k}+m_{k})}\prod_{k=1}^{K}r_{k}^{\alpha_{k}+m_{k}-1} d\textbf{r}\nonumber \\
&=&\frac{\Gamma(\sum_{k=1}^{K}\alpha_{k})}{\Gamma(\sum_{k=1}^{K}(\alpha_{k}+m_{k}))}\prod_{k=1}^{K}\frac{\Gamma(\alpha_{k}+m_{k})}{\Gamma(\alpha_{k})}
\end{eqnarray}

\section{Collapse $\pi_{kl}$ in latent block model with weighted edges}
The collapsed posterior of the latent block model is described in the work by \citep*{Wyse2012}, but the details of the collapsing procedure are not described there. We elaborate the collapsing procedure of the Gaussian latent block model. We collapse $\mu_{kl}$ and $\sigma_{kl}^2$ respectively to get the integral.
\begin{eqnarray}
\int p(\textbf{x}_{kl},\pi_{kl}\vert \textbf{z})d\pi_{kl}&=&\int\int p(\textbf{x}_{kl},\mu_{kl},\sigma_{kl}^2\vert \textbf{z})d\mu_{kl}d\sigma_{kl}^2\nonumber \\
&=&\int\int p(\mu_{kl})p(\sigma_{kl}^2)p(\textbf{x}_{kl}\vert \mu_{kl},\sigma_{kl}^2,\textbf{z})d\mu_{kl}d\sigma_{kl}^2
\end{eqnarray}
To facilitate integrating with respect to $\mu_{kl}$, we denote
\begin{equation}
I_{\mu_{kl}}=\int p(\textbf{x}_{kl},\mu_{kl},\sigma_{kl}^2\vert \textbf{z})d\mu_{kl},
\end{equation}
then
\begin{eqnarray}
I_{\mu_{kl}}
&=&\frac{(\rho/2)^{\nu/2}}{\Gamma(\nu/2)}(2\pi\kappa^2)^{-1/2}(2\pi)^{-w_{kl}/2}\sigma_{kl}^{-1}\sigma_{kl}^{-\nu-2-w_{kl}}\nonumber \\
&&\times\int\mbox{exp}\lbrace -\frac{1}{2\sigma_{kl}^2}[(\frac{1}{\kappa^2}+w_{kl})\mu_{kl}^2-2(\frac{1}{\kappa^2}\xi+s_{kl})\mu_{kl}\nonumber \\
&&+\frac{1}{\kappa^2}\xi^2+q_{kl}+\rho]\rbrace du_{kl}.
\end{eqnarray}
Let 
\begin{equation}
M=\frac{(\rho/2)^{\nu/2}}{\Gamma(\nu/2)}(2\pi\kappa^2)^{-1/2}(2\pi)^{-w_{kl}/2}\sigma_{kl}^{-1}\sigma_{kl}^{-\nu-2-w_{kl}},
\end{equation}
so that 
\begin{eqnarray}
I_{\mu_{kl}}=M \times\int\mbox{exp}\lbrace -\frac{1}{2\sigma_{kl}^2}[\lambda(\mu_{kl}-m)^2-\lambda m^{2} + \frac{1}{\kappa^2}\xi^2+q_{kl}+\rho]\rbrace du_{kl},
\end{eqnarray}
where
\begin{equation}
\lambda=\frac{1}{\kappa^{2}}+w_{kl},
\end{equation}
and
\begin{equation}
m=\frac{\frac{1}{\kappa^{2}}\xi+s_{kl}}{\frac{1}{\kappa^{2}}+w_{kl}}.
\end{equation}
Then
\begin{eqnarray}
I_{\mu_{kl}}&=&M \times (2\pi \frac{\sigma_{kl}^{2}}{\lambda})^{1/2}\int (2\pi \frac{\sigma_{kl}^{2}}{\lambda})^{-1/2}\mbox{exp}\lbrace -\frac{1}{2\sigma_{kl}^2}\lambda(\mu_{kl}-m)^2 \rbrace\nonumber \\
&&\times\mbox{exp}\lbrace -\frac{1}{2\sigma_{kl}^2}(-\lambda m^{2} + \frac{1}{\kappa^2}\xi^2+q_{kl}+\rho)\rbrace du_{kl}\nonumber \\
&=&M \times (2\pi\frac{\sigma_{kl}^{2}}{\lambda})^{1/2}\times\mbox{exp}\lbrace -\frac{1}{2\sigma_{kl}^2}(-\lambda m^{2} + \frac{1}{\kappa^2}\xi^2+q_{kl}+\rho)\rbrace\nonumber \\
&=&(2\pi)^{-w_{kl}/2}\frac{(\rho/2)^{\nu/2}}{\Gamma(\nu/2)}\sigma_{kl}^{-\nu-w_{kl}-2}(w_{kl}\kappa^{2}+1)^{-1/2}\nonumber \\
&&\times\mbox{exp}\lbrace-\frac{1}{2\sigma_{kl}^2}[-\frac{(\frac{1}{\kappa^{2}}\xi+s_{kl})^{2}}{\frac{1}{\kappa^{2}}+w_{kl}}+\frac{1}{\kappa^2}\xi^2+q_{kl}+\rho]\rbrace.
\end{eqnarray}
To facilitate integration with respect to $\sigma_{kl}^{2}$, we first rewrite $I_{\mu_{kl}}$ as follows
\begin{equation}
I_{\mu_{kl}}=(2\pi)^{-w_{kl}/2}\frac{(\rho/2)^{\nu/2}}{\Gamma(\nu/2)}(w_{kl}\kappa^{2}+1)^{-1/2}\frac{\Gamma(\alpha)}{\beta^{\alpha}}\frac{\beta^{\alpha}}{\Gamma(\alpha)}(\sigma_{kl}^{2})^{-(\alpha+1)}e^{(\frac{-\beta}{\sigma_{kl}^{2}})},
\end{equation}
where
\begin{equation}
\alpha=\frac{1}{2}\nu+\frac{1}{2}w_{kl},
\end{equation}
and
\begin{equation}
\beta=\frac{1}{2}[-\frac{(\frac{1}{\kappa^{2}}\xi+s_{kl})^{2}}{\frac{1}{\kappa^{2}}+w_{kl}}+\frac{1}{\kappa^2}\xi^2+q_{kl}+\rho].
\end{equation}
This can be integrated as follows
\begin{eqnarray}
\int I_{\mu_{kl}}d\sigma_{kl}^{2}&=&(2\pi)^{-w_{kl}/2}\frac{(\rho/2)^{\nu/2}}{\Gamma(\nu/2)}(w_{kl}\kappa^{2}+1)^{-1/2}\frac{\Gamma(\alpha)}{\beta^{\alpha}}\nonumber \\
&=&(2\pi)^{-w_{kl}/2}\frac{(\rho/2)^{\nu/2}}{\Gamma(\nu/2)}(w_{kl}\kappa^{2}+1)^{-1/2}\nonumber \\
&&\times \frac{\Gamma(\frac{1}{2}\nu+\frac{1}{2}w_{kl})}{(\frac{1}{2}[-\frac{(\frac{1}{\kappa^{2}}\xi+s_{kl})^{2}}{\frac{1}{\kappa^{2}}+w_{kl}}+\frac{1}{\kappa^2}\xi^2+q_{kl}+\rho])^{(\frac{1}{2}\nu+\frac{1}{2}w_{kl})}}\nonumber \\
&=&\frac{\rho^{\nu/2}\Gamma\lbrace(w_{kl}+\nu)/2\rbrace}{\pi^{w_{kl}/2}\Gamma(\nu/2)(w_{kl}\kappa^2+1)^{1/2}}\nonumber \\
&&\times(-\frac{\kappa^2(s_{kl}+\xi/\kappa^2)^2}{w_{kl}\kappa^2+1}+\frac{\xi^2}{\kappa^2}+q_{kl}+\rho)^{-(w_{kl}+\nu)/2}.
\end{eqnarray}
In summary,
\begin{eqnarray}
\int p(\textbf{x}_{kl},\pi_{kl}\vert \textbf{z})d\pi_{kl}&=&\int\int p(\textbf{x}_{kl},\mu_{kl},\sigma_{kl}^2\vert \textbf{z})d\mu_{kl}d\sigma_{kl}^2\nonumber \\
&=&\frac{\rho^{\nu/2}\Gamma\lbrace(w_{kl}+\nu)/2\rbrace}{\pi^{w_{kl}/2}\Gamma(\nu/2)(w_{kl}\kappa^2+1)^{1/2}}\nonumber \\
&&\times(-\frac{\kappa^2(s_{kl}+\xi/\kappa^2)^2}{w_{kl}\kappa^2+1}+\frac{\xi^2}{\kappa^2}+q_{kl}+\rho)^{-(w_{kl}+\nu)/2}.
\end{eqnarray}

\section{The M3 move} 
In a Gibbs move, only one entry in $\textbf{z}$ is updated at each iteration. An alternative is the M3 move \citep*{Nobile2007}, which updates multiple entries of $\textbf{z}$ simultaneously. In M3, two communities in $\textbf{z}$ are randomly selected and denoted as $k_{1}$ and $k_{2}$. Each element $z_{i}$ in the selected communities is reassigned to $k_{1}$ or $k_{2}$ with probability $P_{k_{1}}^{i}$ and $P_{k_{2}}^{i}$ respectively, to form the updated $\textbf{z}^{\ast}$. The collection of elements of $\textbf{z}$ with labels $k_{1}$ or $k_{2}$ may be indexed by the set $I=\lbrace i : z_{i}=k_{1} \mbox{\ or\ }z_{i}=k_{2}\rbrace$. Let the number of such elements be $L$. The remaining elements of $\textbf{z}$ are collected into a subvector denoted  $\widetilde{\textbf{z}}$. For the update, one element $z_{i}$ with $i \in I$ is randomly selected and updated to $z_{i}^{\ast}$ according to a reassignment probability. The updated element is added to $\widetilde{\textbf{z}}$. The size of $I$ thus becomes $L-1$. This procedure is repeated until all the elements of $I$ are processed (the length of $I$ becomes 0) and the resulting vector $\widetilde{\textbf{z}}$ becomes the proposed move $\textbf{z}^{\ast}$. We define a sub-adjacency matrix $\widetilde{\textbf{x}}$ as the observations corresponding to $\widetilde{\textbf{z}}$ and the observations $\textbf{x}^{{\ast}i}$ corresponding to the updated $z_{i}^{\ast}$.
The probabilities of the reassignment satisfy $P_{k_{1}}^{i}+P_{k_{2}}^{i}=1$ and the ratio
\begin{eqnarray}
\frac{P_{k_{1}}^{i}}{P_{k_{2}}^{i}}&=&\frac{p(z_{i}^{\ast}=k_{1}\vert\widetilde{\textbf{z}},\widetilde{\textbf{x}},\textbf{x}^{\ast i},K)}{p(z_{i}^{\ast}=k_{2}\vert\widetilde{\textbf{z}},\widetilde{\textbf{x}},\textbf{x}^{\ast i},K)}\nonumber \\
&=&\frac{p(z_{i}^{\ast}=k_{1},\widetilde{\textbf{z}},\widetilde{\textbf{x}},\textbf{x}^{\ast i}\vert K)}{p(z_{i}^{\ast}=k_{2},\widetilde{\textbf{z}},\widetilde{\textbf{x}},\textbf{x}^{\ast i}\vert K)}\nonumber \\
&=&\frac{p(z_{i}^{\ast}=k_{1},\widetilde{\textbf{z}}\vert K)}{p(z_{i}^{\ast}=k_{2},\widetilde{\textbf{z}}\vert K)}\frac{p(\widetilde{\textbf{x}},\textbf{x}^{\ast i}\vert z_{i}^{\ast}=k_{1},\widetilde{\textbf{z}},K)}{p(\widetilde{\textbf{x}},\textbf{x}^{\ast i}\vert z_{i}^{\ast}=k_{2},\widetilde{\textbf{z}},K)}.
\end{eqnarray}
The first term of this ratio is given by
\begin{eqnarray}
\frac{p(z_{i}^{\ast}=k_{1},\widetilde{\textbf{z}}\vert K)}{p(z_{i}^{\ast}=k_{2},\widetilde{\textbf{z}}\vert K)}&=&\frac{\Gamma(\alpha_{k_{1}}+\widetilde{m}_{k_{1}}(\widetilde{\textbf{z}})+1)}{\Gamma(\alpha_{k_{1}}+\widetilde{m}_{k_{1}}(\widetilde{\textbf{z}}))}\frac{\Gamma(\alpha_{k_{2}}+\widetilde{m}_{k_{2}}(\widetilde{\textbf{z}}))}{\Gamma(\alpha_{k_{2}}+\widetilde{m}_{k_{2}}(\widetilde{\textbf{z}})+1)}\nonumber \\
&=&\frac{\alpha_{k_{1}}+\widetilde{m}_{k_{1}}(\widetilde{\textbf{z}})}{\alpha_{k_{2}}+\widetilde{m}_{k_{2}}(\widetilde{\textbf{z}})},
\end{eqnarray}
where $\widetilde{m}_{k_{1}}(\widetilde{\textbf{z}})$ and $\widetilde{m}_{k_{2}}(\widetilde{\textbf{z}})$ are the numbers of nodes in community $k_{1}$ and $k_{2}$ in $\widetilde{\textbf{z}}$.
The second term of the ratio is given by
\begin{eqnarray}
\frac{p(\widetilde{\textbf{x}},\textbf{x}^{\ast i}\vert z_{i}^{\ast}=k_{1},\widetilde{\textbf{z}},K)}{p(\widetilde{\textbf{x}},\textbf{x}^{\ast i}\vert z_{i}^{\ast}=k_{2},\widetilde{\textbf{z}},K)}&=&\frac{p(\textbf{x}^{\ast i}\vert\widetilde{\textbf{x}}, z_{i}^{\ast}=k_{1},\widetilde{\textbf{z}},K)}{p(\textbf{x}^{\ast i}\vert\widetilde{\textbf{x}},z_{i}^{\ast}=k_{2},\widetilde{\textbf{z}},K)}\nonumber \\
&=&\frac{p(\textbf{x}^{\ast i}\vert\widetilde{\textbf{x}}^{k_{1}}, z_{i}^{\ast}=k_{1},\widetilde{\textbf{z}},K)}{p(\textbf{x}^{\ast i}\vert\widetilde{\textbf{x}}^{k_{2}},z_{i}^{\ast}=k_{2},\widetilde{\textbf{z}},K)}\nonumber \\
&=&\frac{p(\widetilde{\textbf{x}}^{k_{1}},\textbf{x}^{\ast i}\vert z_{i}^{\ast}=k_{1},\widetilde{\textbf{z}})}{p(\widetilde{\textbf{x}}^{k_{1}}\vert z_{i}^{\ast}=k_{1},\widetilde{\textbf{z}})}\frac{p(\widetilde{\textbf{x}}^{k_{2}}\vert z_{i}^{\ast}=k_{2},\widetilde{\textbf{z}})}{p(\widetilde{\textbf{x}}^{k_{2}},\textbf{x}^{\ast i}\vert z_{i}^{\ast}=k_{1},\widetilde{\textbf{z}})}.
\end{eqnarray}
Finally, the reassignment probability is given by
\begin{eqnarray}
\frac{P_{k_{1}}^{i}}{1-P_{k_{1}}^{i}}&=&\frac{\alpha_{k_{1}}+\widetilde{m}_{k_{1}}(\widetilde{\textbf{z}})}{\alpha_{k_{2}}+\widetilde{m}_{k_{2}}(\widetilde{\textbf{z}})}\frac{p(\widetilde{\textbf{x}}^{k_{1}},\textbf{x}^{\ast i}\vert z_{i}^{\ast}=k_{1},\widetilde{\textbf{z}})}{p(\widetilde{\textbf{x}}^{k_{1}}\vert z_{i}^{\ast}=k_{1},\widetilde{\textbf{z}})}\frac{p(\widetilde{\textbf{x}}^{k_{2}}\vert z_{i}^{\ast}=k_{2},\widetilde{\textbf{z}})}{p(\widetilde{\textbf{x}}^{k_{2}},\textbf{x}^{\ast i}\vert z_{i}^{\ast}=k_{1},\widetilde{\textbf{z}})}.
\end{eqnarray}
and the proposal ratio is given by
\begin{eqnarray}
\frac{p(\textbf{z}^{\ast}\rightarrow\textbf{z})}{p(\textbf{z}\rightarrow\textbf{z}^{\ast})}&=&\prod_{i\in I}\frac{P_{z_{i}}^{i}}{P_{z_{i}^{\ast}}^{i}}.
\end{eqnarray}

\pagebreak

\section{Summary of the algorithm}

\begin{algorithm}
\caption{Bayesian change-point detection by posterior predictive discrepancy}
\hspace*{0.02in} {\bf Input:} 
Time series $\textbf{Y}$ of one subject, length of time course $T$, window size $W$, number of communities $K$.
\begin{algorithmic} [1]

\State For $t = \frac{W}{2}+1,\cdots,T - \frac{W}{2}$
\State \qquad Calculate $\textbf{Y}_{t}\rightarrow \textbf{x}_{t}$.
\State \qquad Draw samples $\lbrace\textbf{z}^{i},\bm{\pi}^{i}\rbrace$ $(i=1,\cdots,S)$ from the posterior $P(\textbf{z},\bm{\pi}\vert\textbf{x},K)$.
\State \qquad Simulate replicated adjacency matrix $\textbf{x}^{rep^{i}}$ from the predictive distribution $P(\textbf{x}^{rep}\vert\textbf{z},\bm{\pi},K)$.
\State \qquad Calculate the disagreement index $\gamma(\textbf{x}^{rep^{i}};\textbf{x})$.
\State \qquad Calculate the posterior predictive discrepancy index $\overline{\gamma}_{t}=\frac{\sum_{i=1}^{S} \bm{\gamma}(\textbf{x}^{rep^{i}};\textbf{x})}{S}$.
\State End
\State For $t =\frac{W}{2}+\frac{W_{s}}{2}+1,\cdots,T-\frac{W}{2}-\frac{W_{s}}{2}$
\State \qquad Calculate cumulative discrepancy energy $E_{t}=\sum_{I=t-\frac{W_{s}}{2}}^{t+\frac{W_{s}}{2}-1}\overline{\gamma}_I$.
\State End
\end{algorithmic}
\end{algorithm}
\section*{Supplementary Table 1:}
\captionsetup[table]{labelfont={bf},name={Supplementary Table},labelsep=period}
\setcounter{table}{0} 
\begin{table}[h!]
\centering
\begin{tabular}{llllllllllllllllll}
\multicolumn{6}{c}{\textbf{2-back}}&\multicolumn{6}{c}{\textbf{0-back}}&\multicolumn{6}{c}{\textbf{Fixation}}\\
Community&\multicolumn{4}{c}{Node number}& &Community&\multicolumn{4}{c}{Node number}& &Community&\multicolumn{4}{c}{Node number}\\
\hline
k=1&29& & & & &k=1& & & & & &k=1& & & & &\\
k=2&11&31&32& & &k=2&31&32& & & &k=2&11&30&31&32&\\
k=3&21& & & & &k=3&16&20&27& & &k=3&12&16&20&21&\\
k=4&1&9&17&34& &k=4&1&9&17&34& &k=4&1&7&9&17&34\\
k=5&2& & & & &k=5&3& & & & &k=5&24& & & &\\
k=6&35& & & & &k=6&5&10& & & &k=6&8& & & &\\
\hline
\end{tabular}
\caption{\footnotesize This table summarises the nodes commonly located in a specific community $k$ for all of picture types in the working memory tasks.}
\label{table:2}
\end{table}

\pagebreak

\setcounter{figure}{0} 
\begin{figure}[!ht]
\centering
\includegraphics[width=0.9\linewidth]{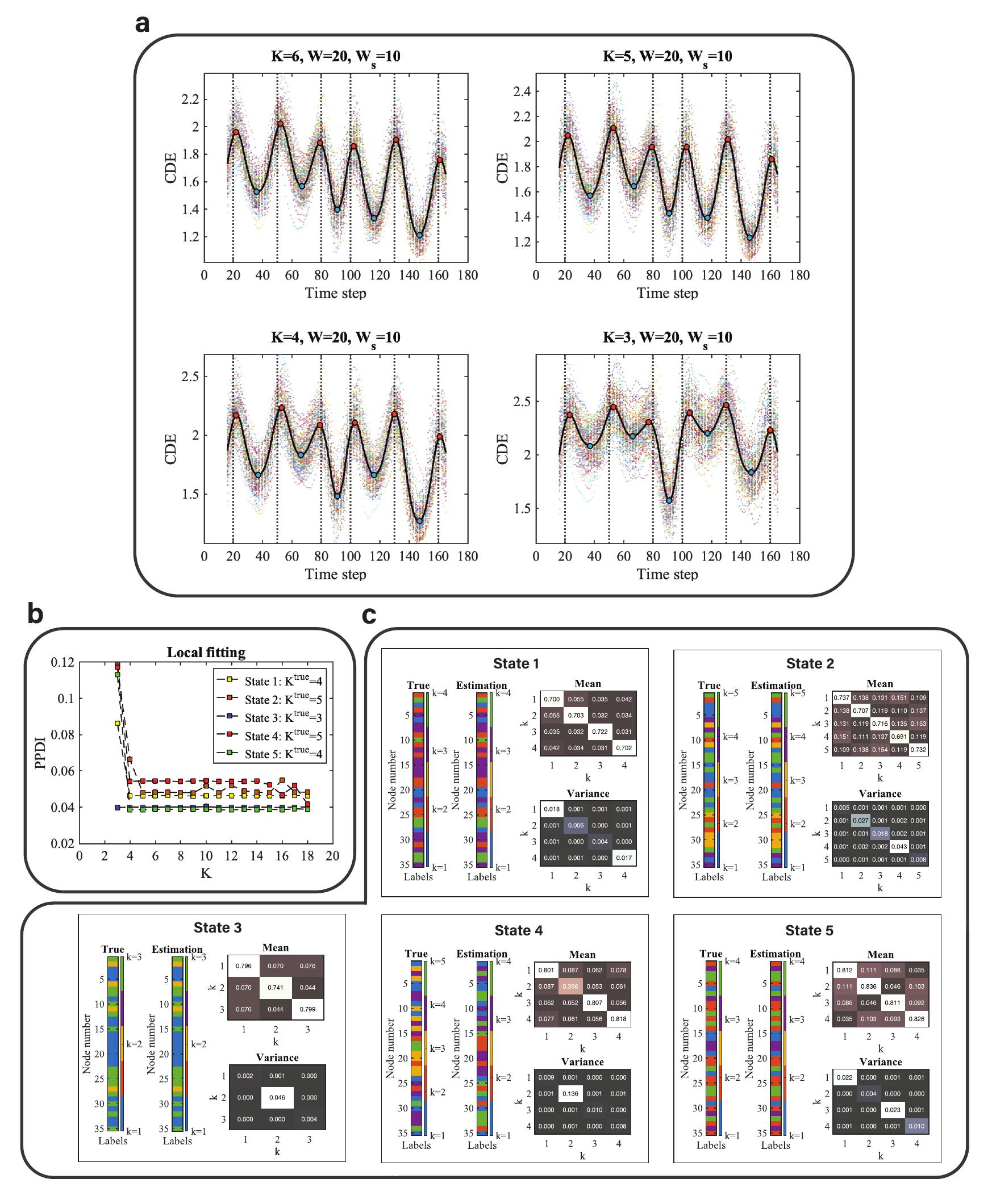}
\caption{\footnotesize \textbf{Results of method validation using synthetic data with SNR = 10dB}: \textbf{a} CDE of the multivariate Gaussian data with SNR = 10dB using different models ($K$=6, 5, 4, and 3). The sliding window size for converting from time series to correlation matrices sequence is $W=20$, whereas (for smoothing) the sliding window size for converting from PPDI to CDE is $W_{s}=10$. The vertical dashed lines are the locations of the true change-points ($t$ = 20, 50, 80, 100, 130, and 160). The colored scatterplots in the figures are the CDEs of individual (virtual) subjects and the black curve is the group CDE (averaged CDE over 100 subjects). The red points are the local maxima and the blue points are the local minima. \textbf{b} Local fitting with different models (from $K$=3 to 18) for synthetic data (SNR=10dB). Different colors represent the PPDI values of different states with the true number of communities $K^{true}$. \textbf{c} The estimation of community constituents for SNR = 10dB at each discrete state: $t$ = 36, 67, 91, 116, 147) for brain states 1 to 5, respectively. The estimations of the latent label vectors (\textbf{Estimation}) and the label vectors (\textbf{True}) that determine the covariance matrix in the generative model are shown as bar graphs. The strength and variation of the connectivity within and between communities are represented by the block mean and variance matrices within each panel.}
\label{synthetic_0_3162}
\end{figure}
\pagebreak

\begin{figure}[!ht]
\centering
\includegraphics[width=0.9\linewidth]{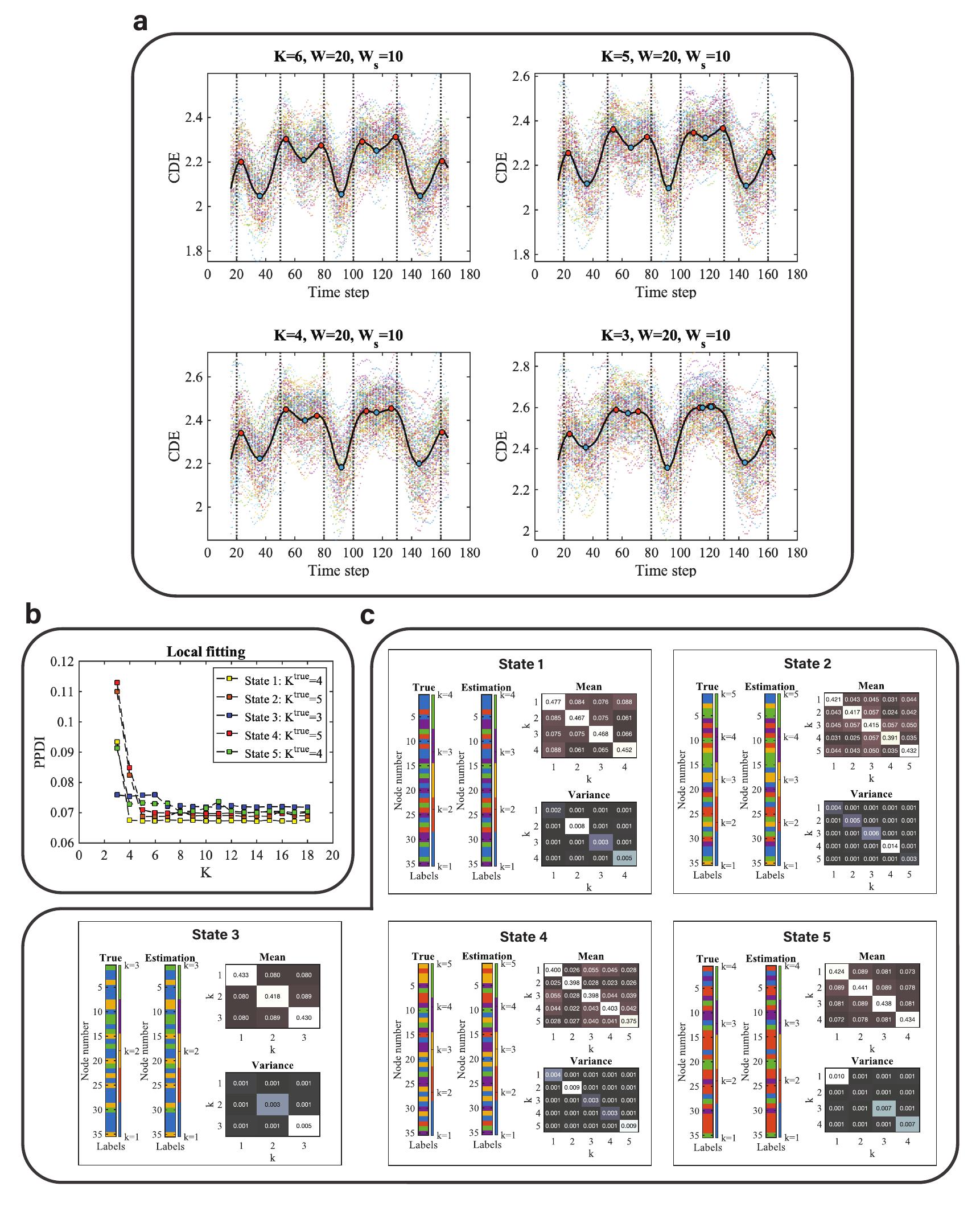}
\caption{\footnotesize \textbf{Results of method validation using synthetic data with SNR = 0dB}: This figure is in the same format as the \textbf{Supplementary Figure 1} above only that it is for \textbf{a} SNR = 0 dB. \textbf{b} Local fitting with different models (from $K$=3 to 18) for synthetic data (SNR=0dB). \textbf{c} The estimation of community constituents for SNR = 0dB at each discrete state: $t$=36, 66, 92, 116, 146) for brain states 1 to 5, respectively.}
\label{synthetic_1}
\end{figure}
\pagebreak

\begin{figure}[!ht]
\centering
\includegraphics[width=0.8\linewidth]{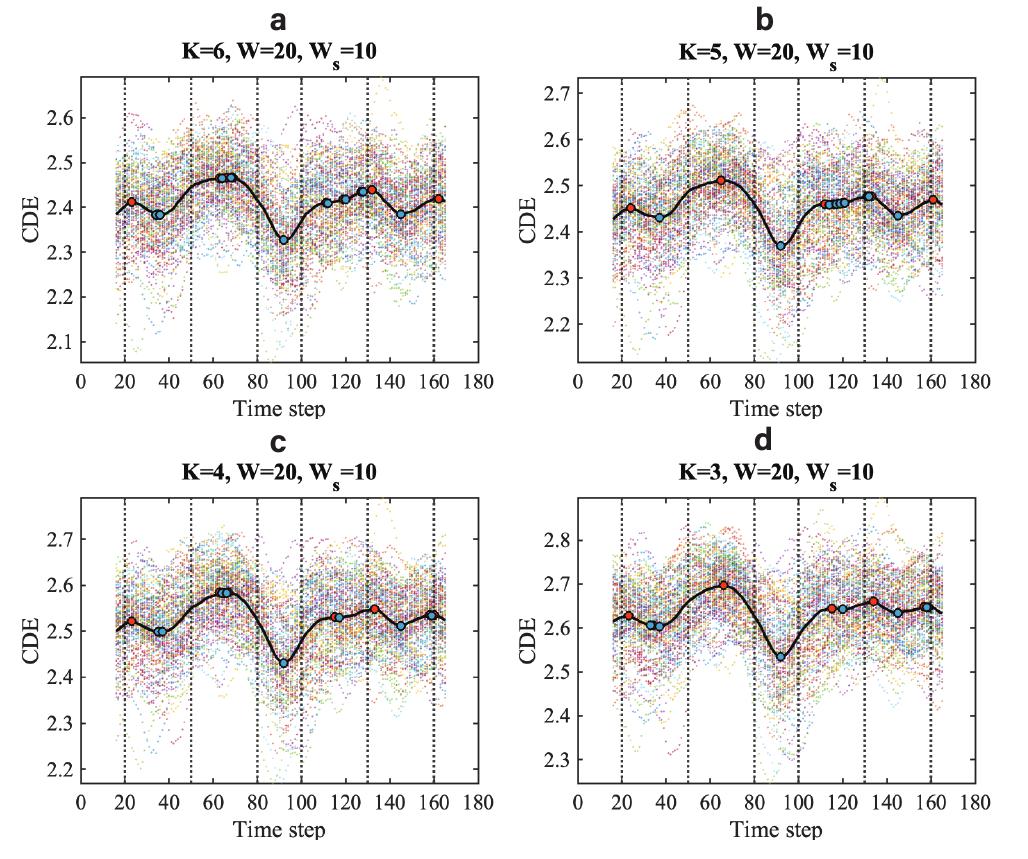}
\caption{\footnotesize CDE of the multivariate Gaussian data with SNR = -5dB using different models ($K$=6, 5, 4, and 3 in \textbf{a} to \textbf{d}). Change-point detection did not work in this case, hence the brain states can not be identified here.}
\label{figure_changepd_sigma1_7783}
\end{figure}
\pagebreak

\begin{figure}[!ht]
\centering
\includegraphics[width=0.65\linewidth]{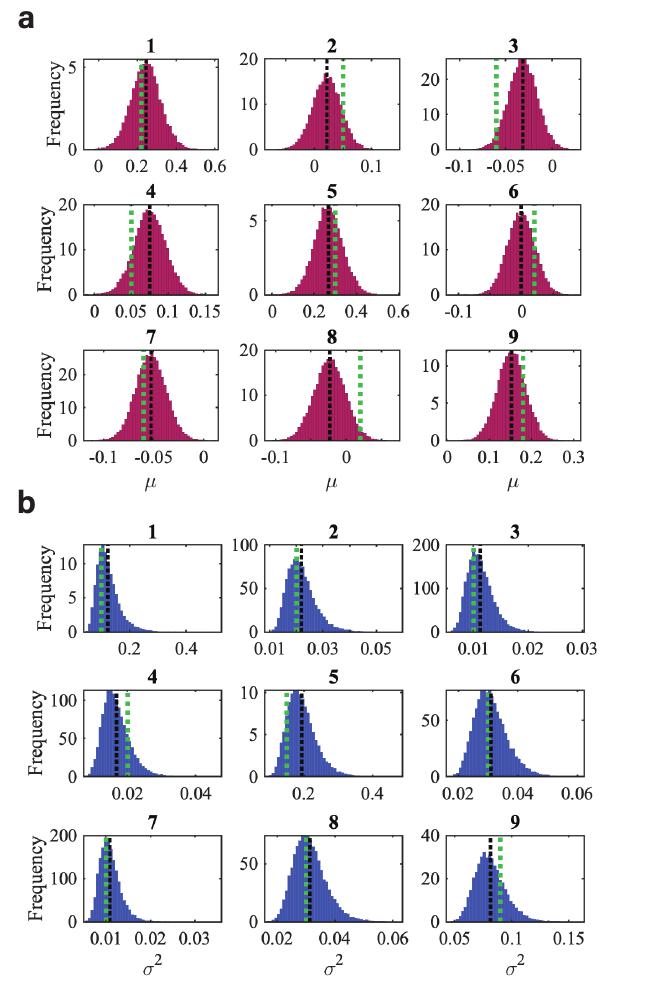}
\caption{\footnotesize \textbf{Validation of sampling the model parameters}: \textbf{a} the histograms of the sampled block mean, \textbf{b} the histograms of the sampled block variance for the case $K=3$. We denote the block $kl$ sequentially (for example, the block for $k=2,l=3$ is denoted as block $6$; the block for $k=3,l=3$ is denoted as block $9$). The green dashed lines are the true values and the black dashed lines are the estimates. In order to validate the algorithm for sampling the model parameters, we simulate a synthetic adjacency matrix from a mixture of Gaussian distributions with ground truth of $K=3$, the true latent label vector (3, 2, 1, 1, 2, 3, 3, 3, 2, 2, 1, 3, 1, 2, 2, 2, 1, 3, 3, 3, 3), the true block mean matrix (0.22, 0.05, -0.06; 0.05, 0.30, 0.02; -0.06, 0.02, 0.18) and the true block variance matrix (0.1, 0.02, 0.01; 0.002, 0.15, 0.03; 0.01, 0.03, 0.09). Given this generated adjacency matrix as an observation, we draw samples of the block mean and variance from the posterior $p(\bm{\pi|\textbf{x},\textbf{z}})$ conditional on $\textbf{z}$. The shape of the histogram of mean is consistent with a Normal distribution and the histogram of variance is consistent with an Inverse-Gamma distribution. The figure shows that the estimations of the block mean and variance closely match the ground truth values.}
\label{Supp_3_syn_meanvari_validation}
\end{figure}
\pagebreak

\begin{figure*}[!ht]
\centering
\includegraphics[width=0.75\linewidth]{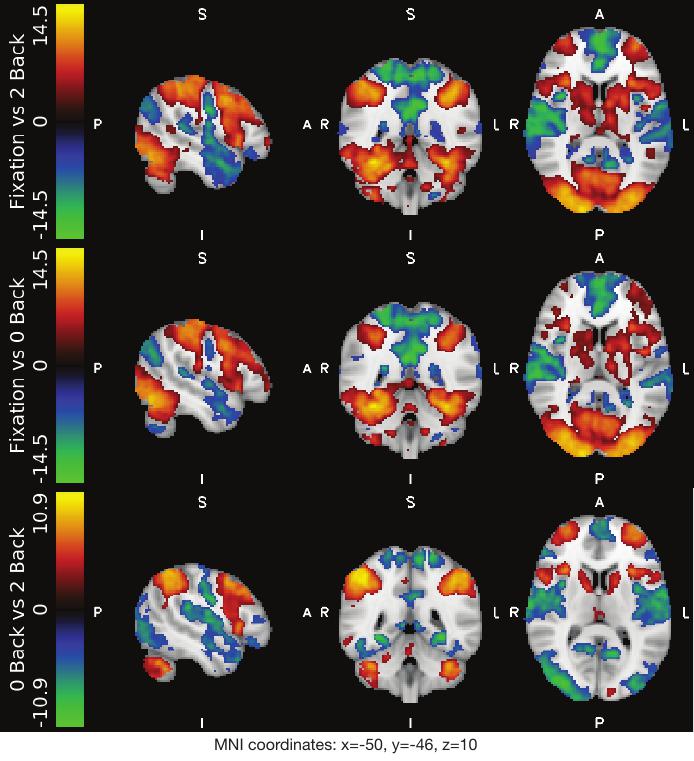}
\caption{\footnotesize \textbf{Task activation maps (thresholded Z-MAX maps) for group analysis}: Contrasts of 2-back vs fixation, 0-back vs fixation and 2-back vs 0-back for MNI coordinates (x = -50, y = -46, z = 10). For running 1st-level GLM-based FEAT \citep*{Woolrich2001} in FSL, we added the confound predictors files released by HCP to the design matrix of the model for each individual. We then set up a 2nd-level design matrix for the contrast of 2-back, 0-back, and fixation. For the 3rd-level (the group-level GLM analysis \citep*{Woolrich2004}), we applied cluster-wise inference and set up the cluster defining threshold (CDT) to be $Z=3.1$ ($P=0.001$) to avoid cluster failure problems as described in \citep*{Eklund2016}, with a family-wise error-corrected threshold of $P=0.05$. Maps are viewed by looking upward from the feet of the subject and the coordinate directions are denoted as Anterior (A), Posterior (P), Superior (S), Inferior (I), Left (L), and Right (R).}
\label{Figure_4_contrasts_WM}
\end{figure*}
\pagebreak

\begin{figure}[!ht]
\centering
\includegraphics[width=0.8\linewidth]{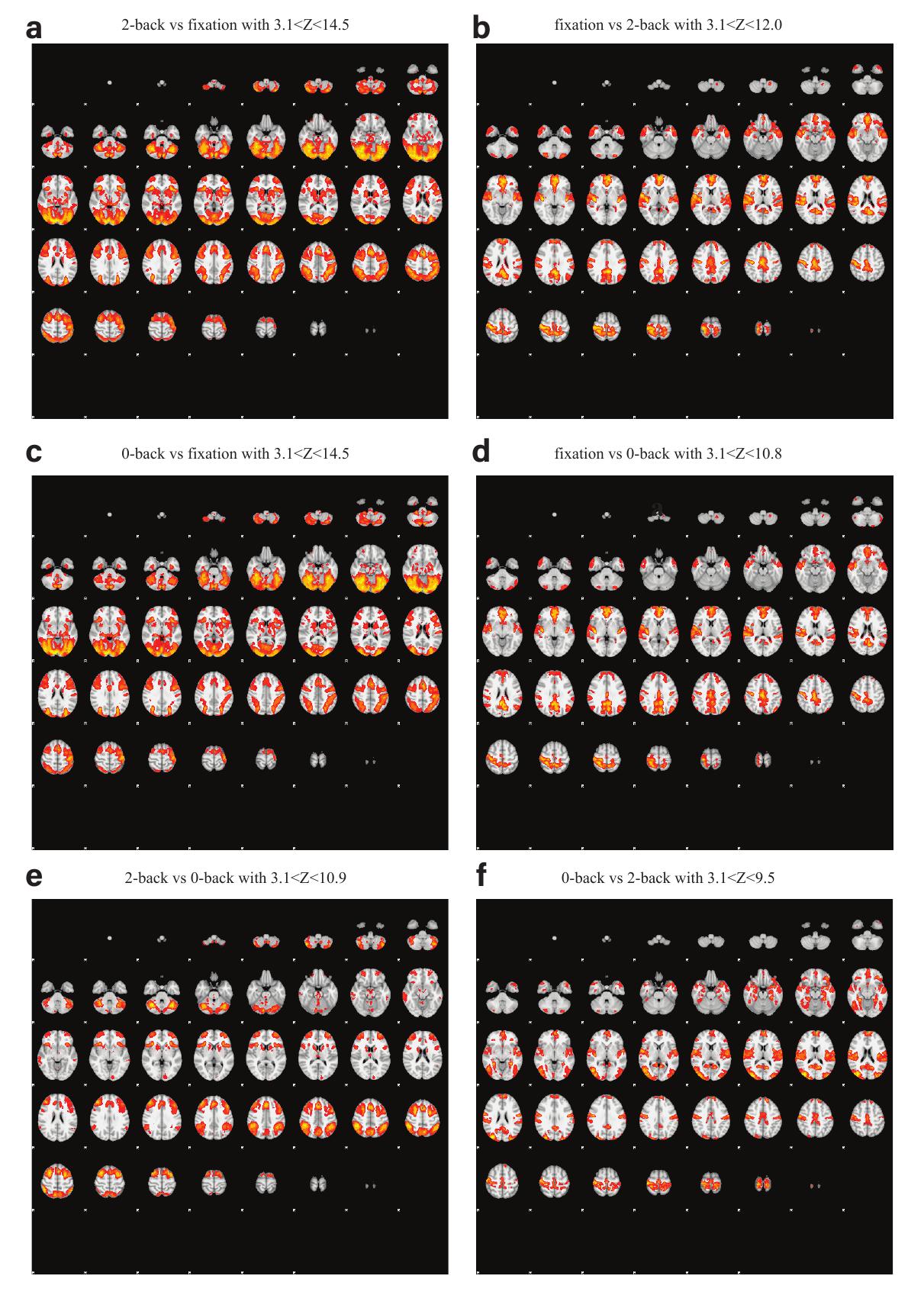}
\caption{\footnotesize \textbf{Thresholded activation maps}}
\label{contrast_lightbox}
\end{figure}

\pagebreak

\begin{figure*}[!ht]
\centering
\includegraphics[width=0.8\linewidth]{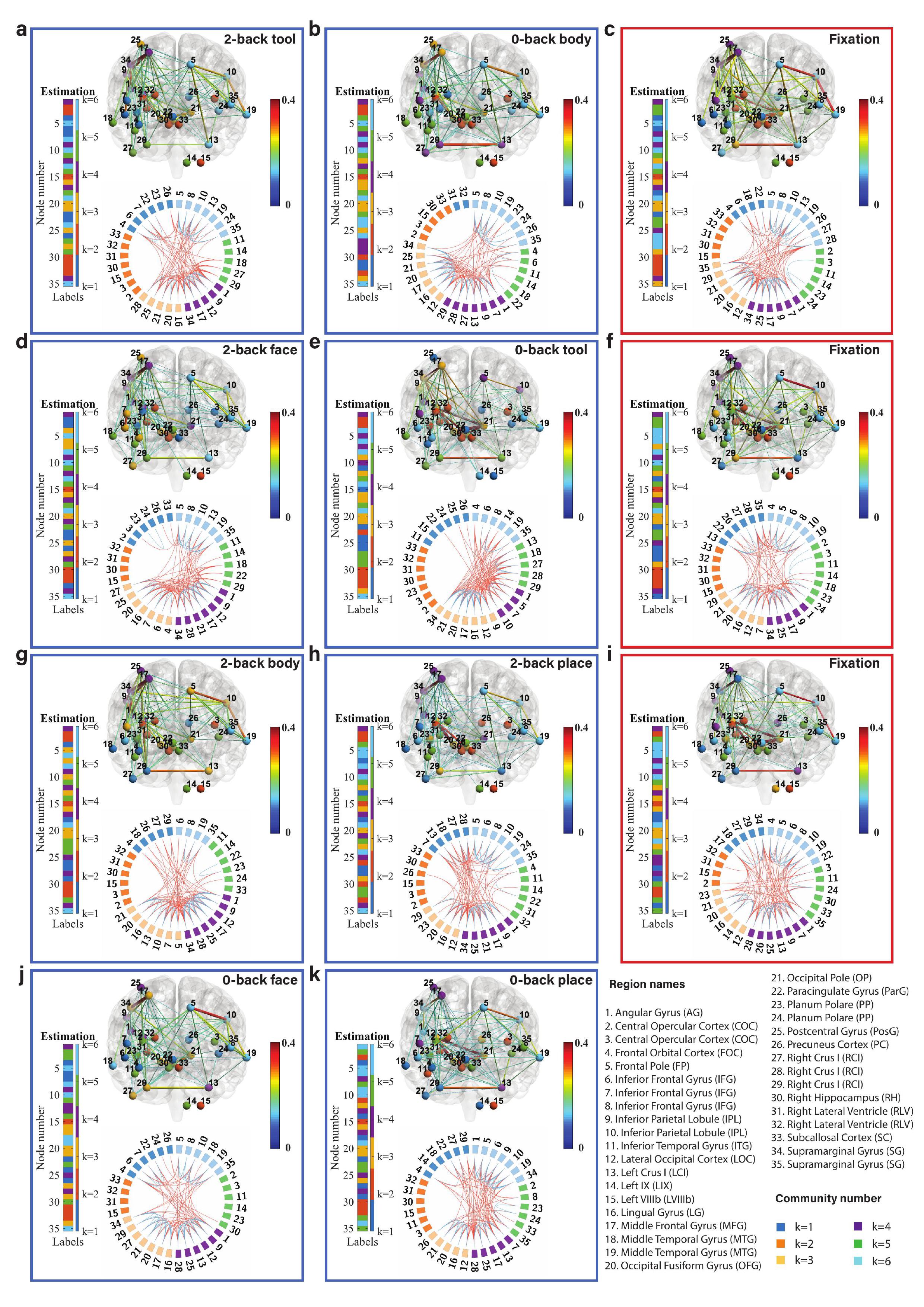}
\caption{\footnotesize \textbf{Community structure of the discrete brain states with sparsity level of 20\% (session 1: LR)}: The figures with blue frames represent brain states corresponding to working memory tasks (2-back tool at $t=41$; 0-back body at $t=76$; 2-back face at $t=140$; 0-back tool at $t=175$; 2-back body at $t=239$; 2-back place at $t=278$; 0-back face at $t=334$; and 0-back place at $t=375$ in \textbf{a}-\textbf{k}) and those with red frames represent brain states corresponding to fixation (fixation at $t=107, 206,$ and $306$ in \textbf{c}, \textbf{f}, and \textbf{i}). Each brain state shows connectivity at a sparsity level of 20\%. The different colors of the labels represent community memberships. The strength of the connectivity is represented by the colors shown in the bar at the right of each frame. In Circos maps, nodes in the same community are adjacent and have the same color. Node numbers and abbreviations of the corresponding brain regions are shown around the circles. In each frame, different colors represent different community numbers. The connectivity above the sparsity level is represented by arcs. The blue links represent connectivity within communities and the red links represent connectivity between communities.
}
\label{brainnet_sparsity_20}
\end{figure*}
\pagebreak

\begin{figure*}[!ht]
\centering
\includegraphics[width=0.8\linewidth]{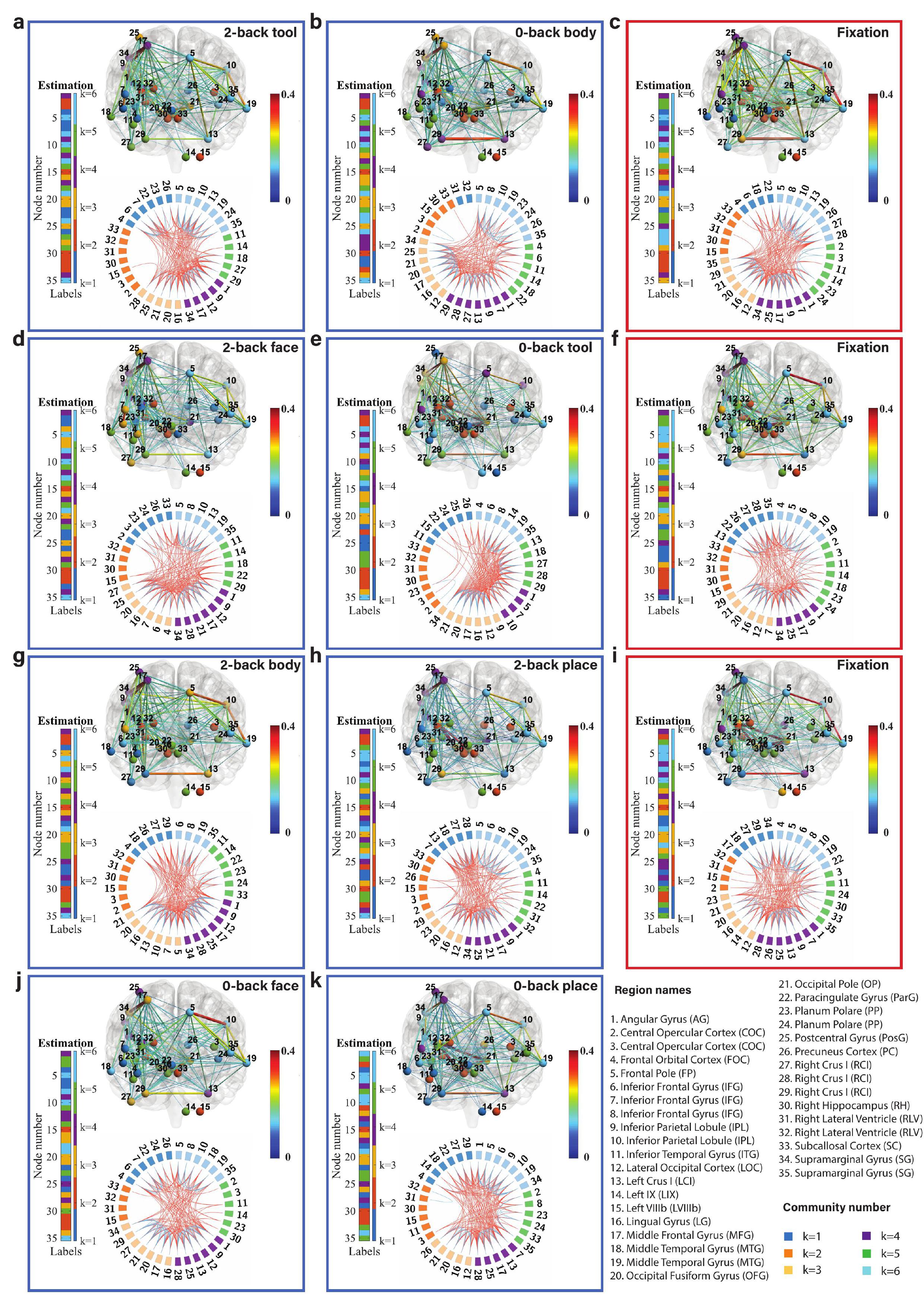}
\caption{\footnotesize \textbf{Community structure of the discrete brain states with sparsity level of 30\% (session 1: LR)}: This figure is in the same format as the \textbf{Supplementary Figure 7} above only that it is for sparsity level of 30\%.
}
\label{brainnet_sparsity_30}
\end{figure*}
\pagebreak

\begin{figure*}[!ht]
\centering
\includegraphics[width=0.8\linewidth]{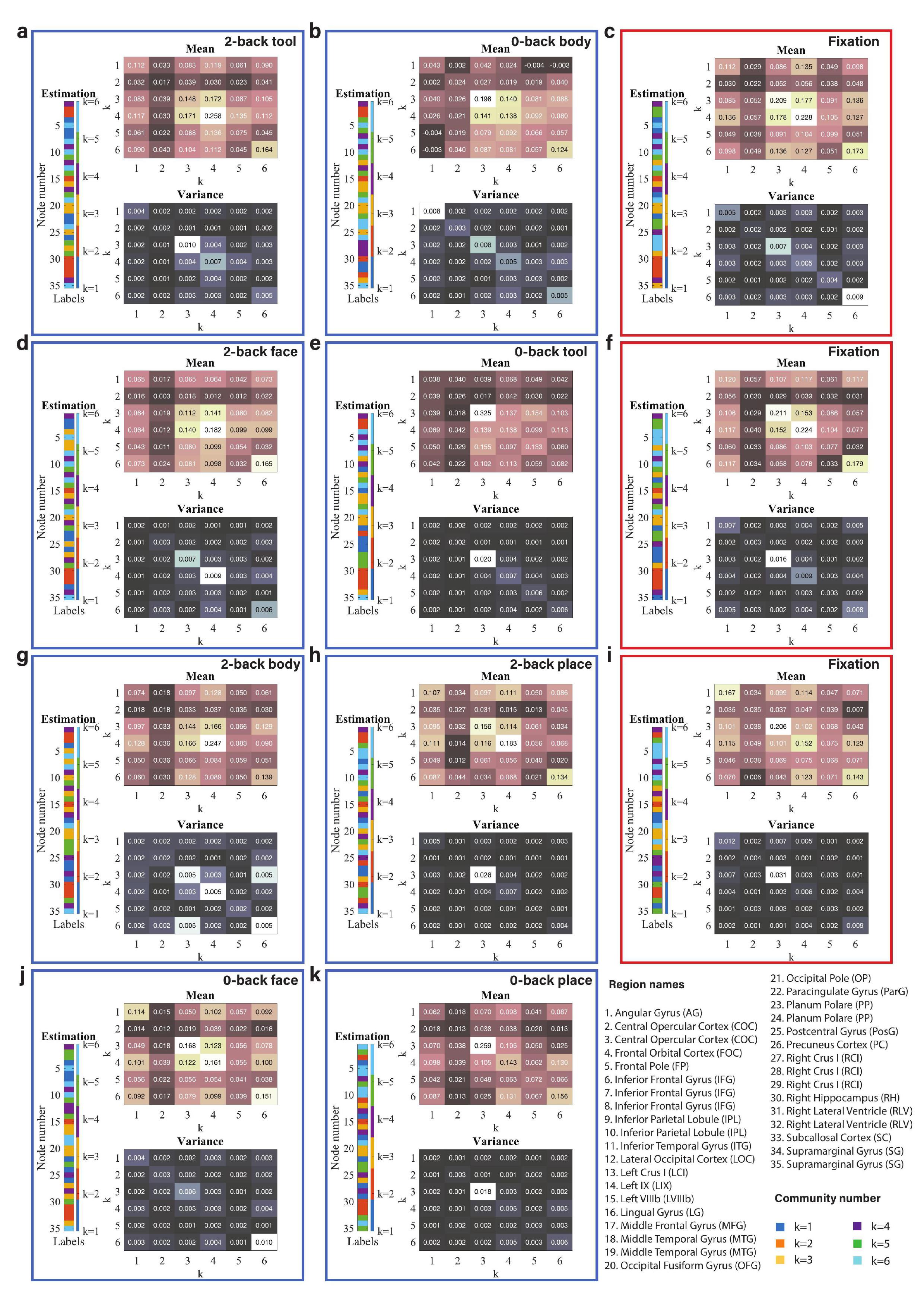}
\caption{\footnotesize \textbf{Estimated mean and variance matrices of the blocks for brain states (session 1: LR)}: The figures with blue frames represent brain states corresponding to working memory tasks (2-back tool at $t=41$; 0-back body at $t=76$; 2-back face at $t=140$; 0-back tool at $t=175$; 2-back body at $t=239$; 2-back place at $t=278$; 0-back face at $t=334$; and 0-back place at $t=375$ in \textbf{a}-\textbf{k}) and those with red frames represent brain states corresponding to fixation (fixation at $t=107, 206,$ and $306$ in \textbf{c}, \textbf{f}, and \textbf{i}). The different colors of the labels represent community memberships. The density and variation of connectivity within and between communities are shown in the estimated block mean matrix and block variance matrix.
}
\label{mean_variance_matrix}
\end{figure*}
\pagebreak

\begin{figure*}[!ht]
\centering
\includegraphics[width=0.8\linewidth]{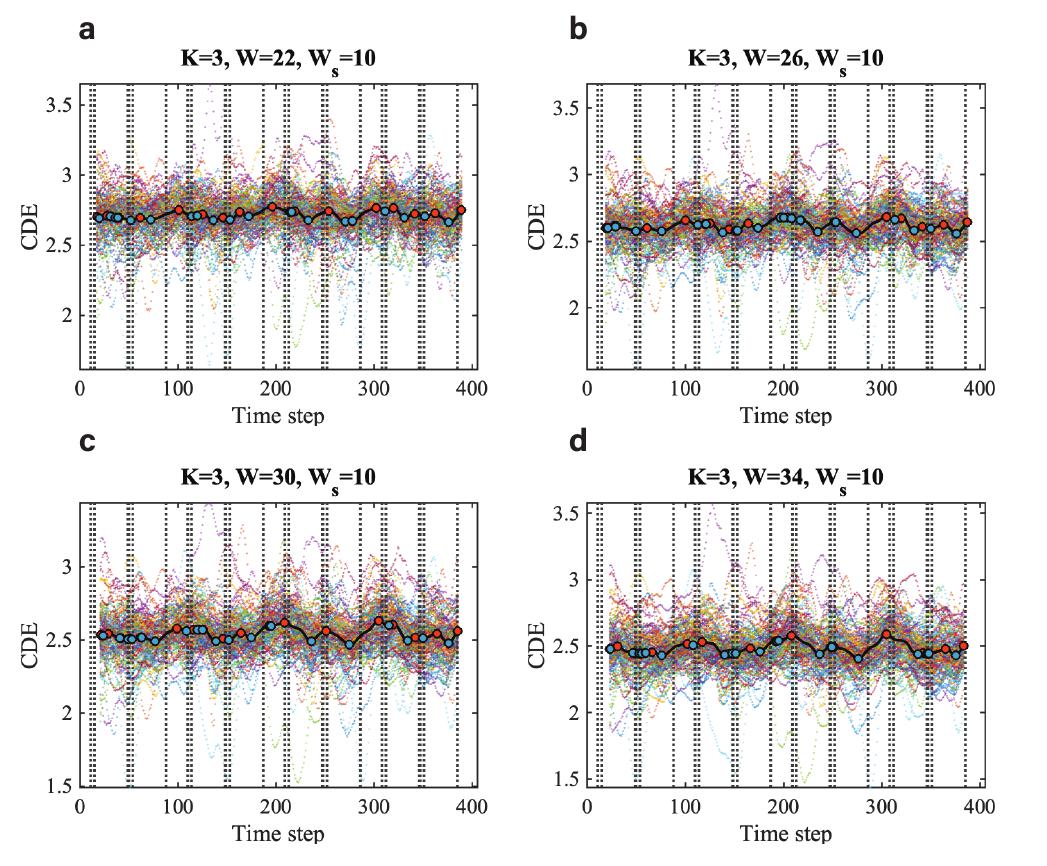}
\caption{\footnotesize \textbf{Results of Bayesian change-point detection for working memory tfMRI data (session 2 RL)}: Cumulative discrepancy energy (CDE) with different sliding window sizes (\textbf{a}  $W=22$, \textbf{b}  $W=26$, \textbf{c} $W=30$ and \textbf{d} $W=34$ under the model $K=3$) and different models (\textbf{c} K=3; \textbf{e} K=4; \textbf{f} K=5 using a sliding window of $W=30$). $W_{s}$ is the sliding window used for converting from PPDI to CDE. The vertical dashed lines are the times of onset of the stimuli, which are provided in the EV.txt files in the released data. The colourful scatterplots in the figures represent the CDEs of individual subjects and the black curve is the group CDE (averaged CDE over 100 subjects). The red points are the local maxima, which are taken to be the locations of change-points, and the blue points are the local minima, which are used for local inference of the discrete brain states.}
\label{Figure_5_changepoint}
\end{figure*}
\pagebreak

\begin{figure}[!ht]
\centering
\includegraphics[width=0.7\linewidth]{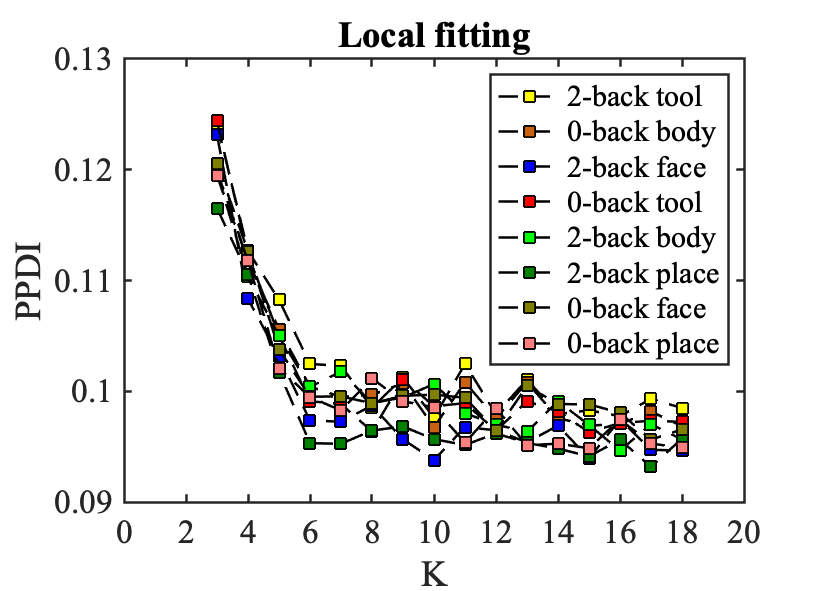}
\caption{\footnotesize Local fitting (session 2 RL) between averaged adjacency matrix and models from $K=3$ to $K=18$. Different colours represent the PPDI values of different brain states.}
\label{localfit_real}
\end{figure}
\pagebreak

\begin{figure*}[!ht]
\centering
\includegraphics[width=0.8\linewidth]{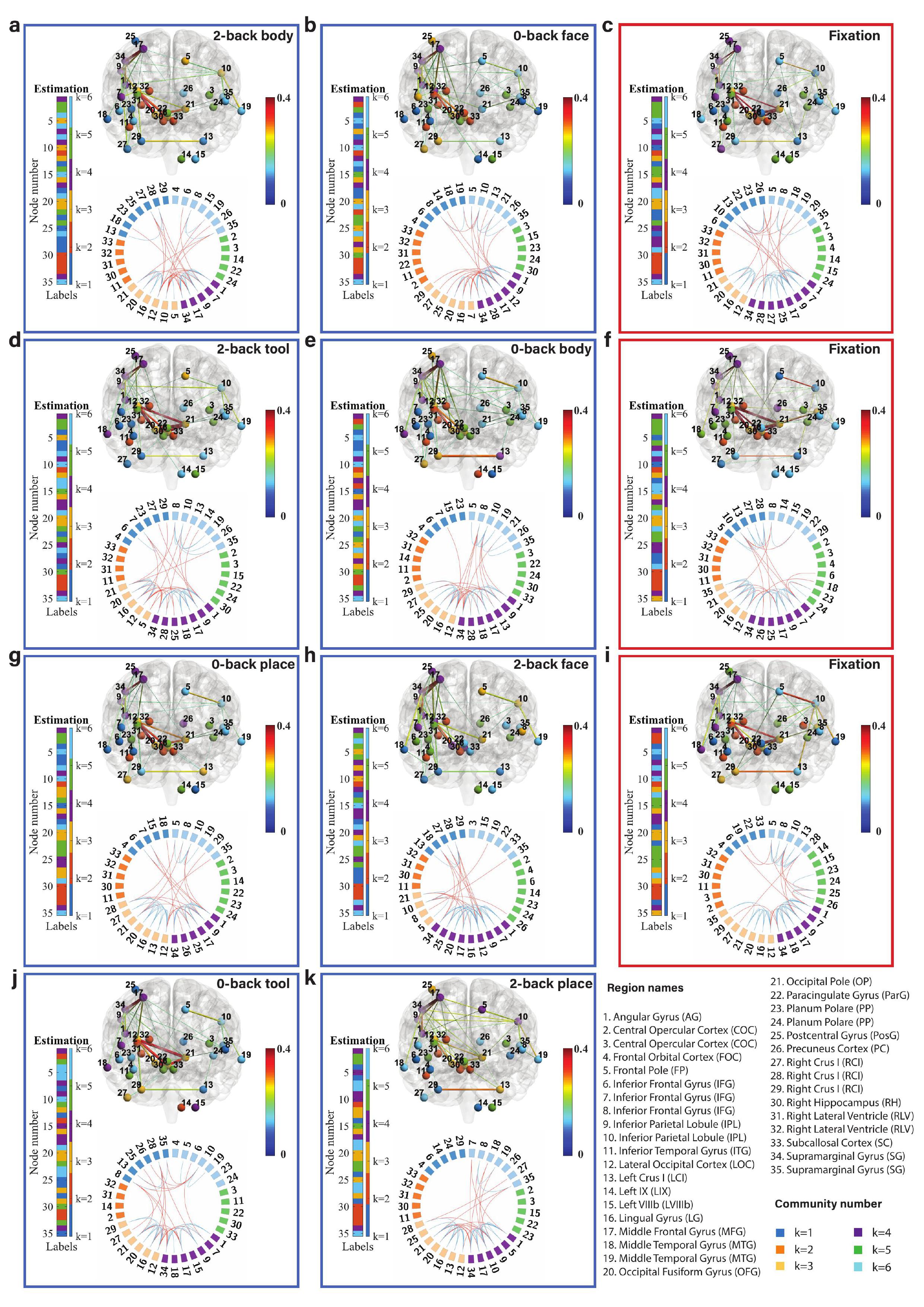}
\caption{\footnotesize \textbf{Community structure of the discrete brain states with sparsity level of 10\% (session 2: RL)}: The figures with blue frames represent brain states corresponding to working memory tasks (2-back tool at $t=41$; 0-back body at $t=77$; 2-back face at $t=139$; 0-back tool at $t=175$; 0-back body at $t=236$; 2-back place at $t=275$; 0-back face at $t=334$; and 2-back place at $t=376$ in \textbf{a}-\textbf{k}) and those with red frames represent brain states corresponding to fixation (fixation at $t$=99, 209, and 306 in \textbf{c}, \textbf{f}, and \textbf{i}).
}
\label{brainnet_sparsity_10}
\end{figure*}
\pagebreak

\begin{figure*}[!ht]
\centering
\includegraphics[width=0.8\linewidth]{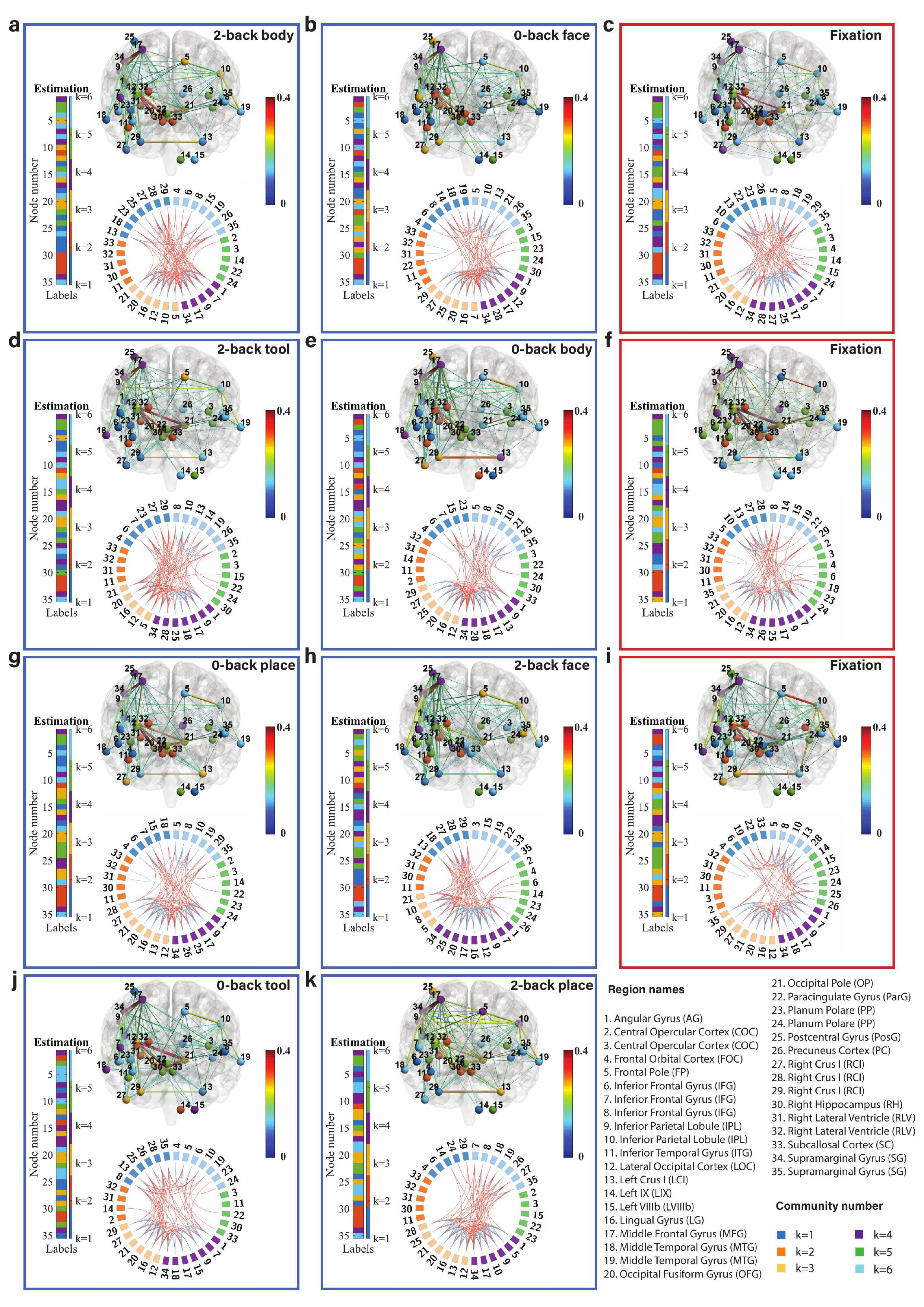}
\caption{\footnotesize \textbf{Community structure of the discrete brain states with sparsity level of 20\% (session 2: RL)}: This figure is in the same format as the \textbf{Supplementary Figure 11} above only that it is for sparsity level of 20\%.
}
\label{brainnet_sparsity_10}
\end{figure*}
\pagebreak

\begin{figure*}[!ht]
\centering
\includegraphics[width=0.8\linewidth]{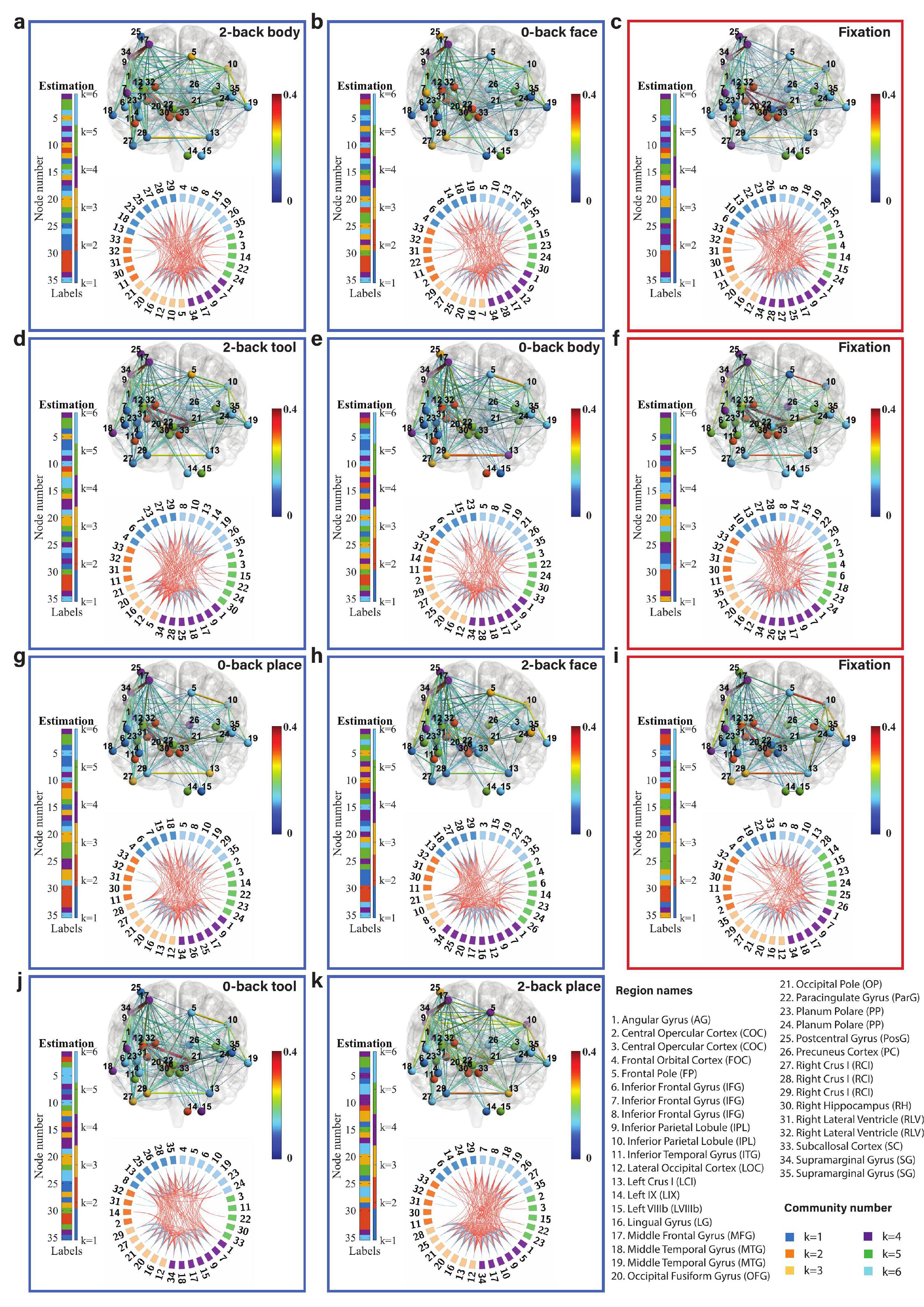}
\caption{\footnotesize \textbf{Community structure of the discrete brain states with sparsity level of 30\% (session 2: RL)}: This figure is in the same format as the \textbf{Supplementary Figure 11} above only that it is for sparsity level of 30\%. 
}
\label{brainnet_sparsity_10}
\end{figure*}
\pagebreak

\begin{figure*}[!ht]
\centering
\includegraphics[width=0.8\linewidth]{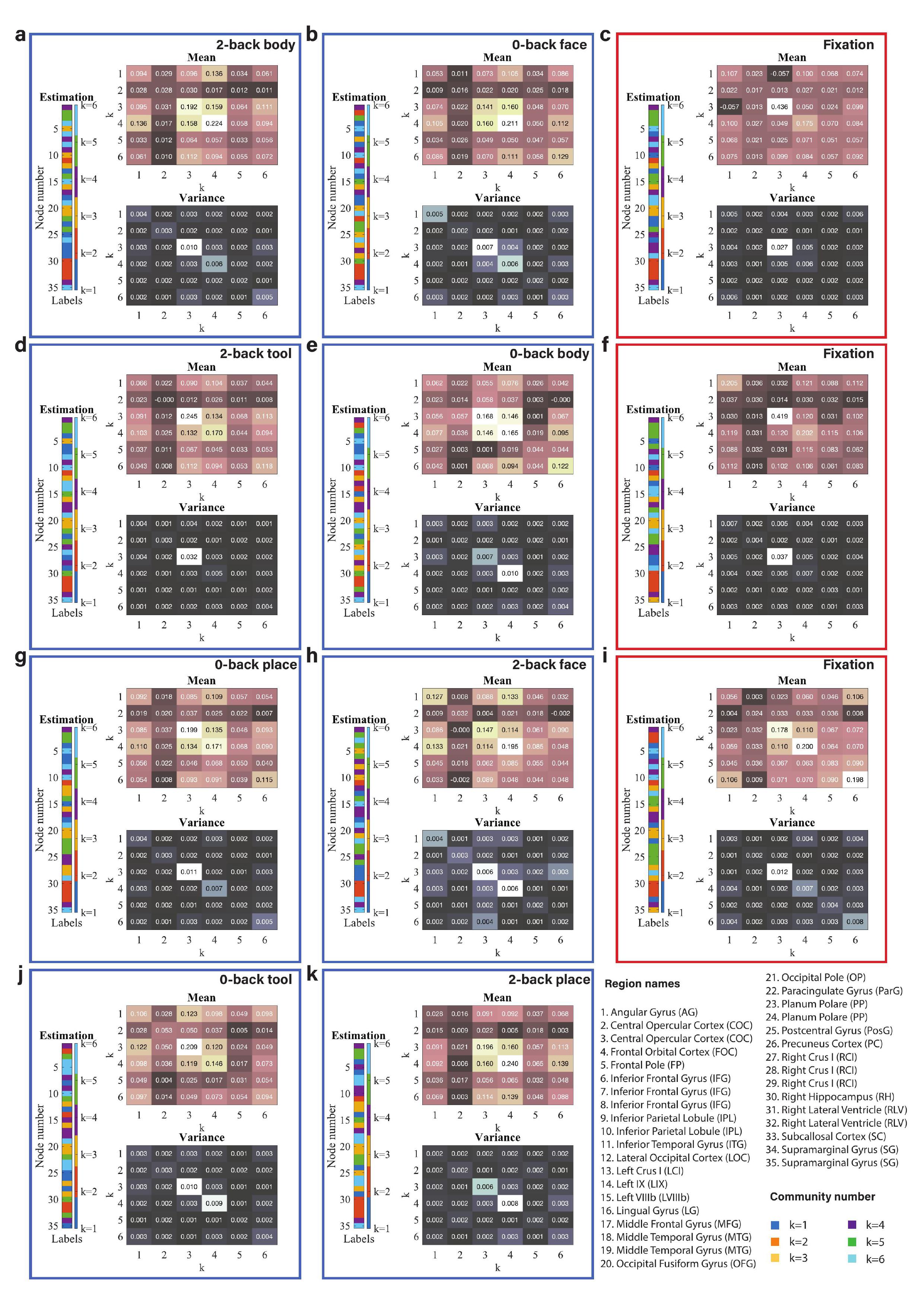}
\caption{\footnotesize \textbf{Estimated mean and variance matrices of the blocks for brain states (session 2: RL)}: This figure is in the same format as the \textbf{Supplementary Figure 9}. The figures with blue frames represent brain states corresponding to working memory tasks (2-back tool at $t=41$; 0-back body at $t=77$; 2-back face at $t=139$; 0-back tool at $t=175$; 0-back body at $t=236$; 2-back place at $t=275$; 0-back face at $t=334$; and 2-back place at $t=365$ in \textbf{a}-\textbf{k}) and those with red frames represent brain states corresponding to fixation (fixation at $t=99, 209,$ and $306$ in \textbf{c}, \textbf{f}, and \textbf{i}).}
\label{mean_variance_matrix}
\end{figure*}
\pagebreak

\clearpage
%

\end{document}